\documentclass[journal,twoside,web]{IEEEtran}

\usepackage{graphicx}
\usepackage{amsmath}
\usepackage[utf8]{inputenc}
\usepackage{color}

\usepackage{graphicx}
\usepackage{rotating}
\usepackage{tikz}
\usepackage{amssymb}
\usepackage{epsfig}
\usepackage{subfigure}
\usepackage{epstopdf}
\usepackage{algorithm}
\usepackage{algorithmic}
\usepackage{amsmath}
\usepackage{amsfonts}
\usepackage{dsfont}
\usepackage{url}
\usepackage{chemarr}
\usepackage{chemarrow}
\usepackage{bbold}
\usepackage[version=3]{mhchem}
\usepackage{mathabx}
\usepackage{tabularx,booktabs}
\usepackage{booktabs}
\usepackage[font=small,skip=0pt]{caption}

\ifCLASSINFOpdf
\else
\fi
\begin{document}
%
\title{Communication and Information Theory of Single Action Potential Signals in Plants}
%
%
%

\author{Hamdan Awan, Raviraj S. Adve, Nigel Wallbridge,\\ Carrol Plummer, and Andrew W. Eckford%
\thanks{Draft: \today}%
\thanks{Hamdan Awan and Andrew W. Eckford are with the Department
of Electrical Engineering and Computer Science, York University, Toronto, Ontario, Canada M3J 1P3. E-mails: hawan@eecs.yorku.ca, aeckford@yorku.ca.}%
\thanks{Raviraj S. Adve is with The Edward S. Rogers Sr. Department of Electrical and Computer Engineering, University of Toronto, Toronto, Ontario, Canada M5S 3G4. E-mail: rsadve@ece.utoronto.ca}%
\thanks{Nigel Wallbridge and Carrol Plummer are with Vivent SaRL, 1299 Crans-pr\`es-C\'eligny, Switzerland. E-mails: nigel.wallbridge@vivent.ch, carrol.plummer@vivent.ch}%
\thanks{Material in this paper is accepted for publication in the 2018 IEEE Global Communications Conference (GLOBECOM).}%
}
%
%

\markboth{IEEE Transactions on NanoBioScience}%
{Accepted paper}
%



\maketitle

\begin{abstract}
Many plants, such as {\em Mimosa pudica} (the ``sensitive plant''), employ electrochemical signals known as action potentials (APs) for rapid intercellular communication.
In this paper, we consider a reaction-diffusion model of individual AP signals to analyze APs from a communication- and information-theoretic perspective.
We use concepts from molecular communication to explain the underlying process of information transfer in a plant for a single AP pulse that is shared with one or more receiver cells. 
We also use the chemical Langevin equation to accommodate the deterministic as well as stochastic component of the system. 
Finally we present an information-theoretic analysis of single action potentials, obtaining achievable information rates for these signals. We show that, in general, the presence of an AP signal can increase the mutual information and information propagation speed among neighboring cells with receivers in different settings.
\end{abstract}


%
\IEEEpeerreviewmaketitle

\section{Introduction}

\label{sec:intro}
Action potentials (APs) are electrochemical signals in biological communication systems. Though commonly associated with the firing of neurons, APs also play a significant role in plants. For example, {\em Mimosa pudica}, the ``sensitive plant'', closes its leaves when touched: the signal to close the leaves is carried by an AP, as proposed by Bose over a century ago \cite{bose1914}. As the plant closes its leaves, it startles herbivorous insects and causes them to leave the {\em Mimosa} alone \cite{pickard1973}.

A plant AP signal can be defined as a sudden change or increase in the resting potential of the cell as a result of some external \textcolor{black}{stimulus} \cite{sukhov2009mathematical}. \textcolor{black}{ The AP in plants differs from the neural AP in its propagation mechanisms. In a neural AP, the signals propagate along the neuron's axon towards synaptic boutons which are present at the ends of an axon. These signals then connect with other neurons at synapses \cite{kress2009action}. However, in a plant AP, the signals propagate from one cell to neighboring cells which are connected through plasmodesmata (a narrow thread of cytoplasm that passes through the cell walls of adjacent plant cells and allows communication between them) \cite{fromm2007electrical}. }

Mathematical models for AP generation in plants are known \cite{sukhov2009mathematical,sukhov2011simulation}, and research on electrical signals and \textcolor{black}{the associated} physiological or biochemical response in plants is an active area of ongoing research \cite{fromm2007electrical,fromm1995biochemical}.
\textcolor{black}{The electrical signals can influence different processes in the plant. An example for this is the effect of electrical signals on photosynthesis which is reviewed in different works such as \cite{sukhov2016electrical,szechynska2017electrical}. Similarly, the impact of electrical signals and plant tolerance was analyzed in  literature \cite{surova2016variation,sukhov2015variation,sukhov2017high}. Also the effect of electrical signals on different functionalities of plant are discussed in the literature, such as respiration \cite{lautner2014involvement}, gene expression \cite{pena1995signals}, hormones production \cite{hlavavckova2006electrical}, ATP content \cite{surova2016variation}, and others. Similarly the mechanisms of electrical signals influence on physiological activity of plants (specifically its cells and environment) are studied in  \cite{sukhova2018influence}. Some mathematical models of electrical signals influence on physiological processes are presented in \cite{sherstneva2015participation}.
}

\textcolor{black} {Some models of AP for different type of plants such as algae are presented in the literature. For example the action potential in {\em Characeae} is presented in \cite{beilby1984current}. This paper used the experimental techniques of high sophistication to find out the role of calcium ions in the generation of the action potential.  This model is further investigated by the authors in \cite{beilby2016re} which provided new insights into the AP in these organisms, and suggested a range of experiments. Another work is presented in \cite{mummert1991action}, which focuses on the action potentials in another type of algae i.e. {\em Acetabularia}. Several other models of electrical signals are reviewed by \cite{sukhova2017mathematical}.}

The AP signal is associated with passive fluxes of ion channels such as calcium, chlorine and potassium  in the cell \cite{fromm2007electrical,fromm1995biochemical,felle2007systemic}. This means that the stationary level of membrane potential (resting potential) is changed by an external \textcolor{black}{stimulus} 
(electrical or environmental) leading to the generation of an AP signal. \textcolor{black}{The different (intracellular and intercellular) electrical signals,} including the AP signal, play a significant role in enabling the plant to adopt to the change in the environment.

\textcolor{black} {The AP signal depends on number of factors, such as the change in the ion concentrations as a result of external stimulus.  A more recent  study suggests that the AP signal also depends on the area and volume of vacuole inside the cell \cite{novikova2017mathematical}. }

\textcolor{black} {
It is clear from the models in \cite{evans2017chemical,vodeneev2018parameters,sukhova2017mathematical} that understanding of plant potentials such as variation potentials is informed by molecular communication \cite{nakano2013-book,farsad2016comprehensive}, a communication paradigm inspired by the communication between living cells \cite{Akyildiz:2008vt,Hiyama:2010jf,Nakano:2014fq}.  Another example of the role  of chemical signals in transmission of signals from cells with electrical responses to other cells is given in \cite{gilroy2016ros}. In this paper, we simulate the AP generation and propagation by means of molecules propagating from one cell to another in different configurations. Based on the existing literature, it is reasonable to assume that chemical signals can also be a mechanism for APs similar to variation potentials. In future work, we will investigate the use of ion-based electrical signals as the mechanism for propagation of APs, in order to gain understanding of the impact of different mechanisms on the communication properties of the overall system.  }

A key characteristic of molecular communication is the use of molecules as the information or signal carrier. The transmission of signalling molecules can be carried out by diffusion \cite{Pierobon:2010kz} or active transport \cite{farsad2011simple}. An earlier paper \cite{Chou:2014jca} considers a few different types of reactions at the receiver, including a linearized form of ligand-receptor binding, catalysis and regulated catalysis. 

The current paper considers the system where the \textcolor{black}{receiver is based on chemical reactions, i.e. a linearized form of ligand-receptor binding similar to previous works \cite{awan2016generalized,awan2016demodulation,Awan:2016:RER:2967446.2967455,Awan:2015:IRM:2800795.2800798,riaz2018using}}. Furthermore, in this paper we aim to compute mutual information between the input number of signaling molecules  (based on action potential signal) and the output number of molecules produced \textcolor{black}{by a number of receiver cells in different configurations. }

Our first aim is to present a physical and mathematical model for the generation of single action potential signals \textcolor{black}{in a plant cell}. The next aim is  to consider the impact of this AP signal on number of output molecules \textcolor{black}{based on} the concept of diffusion-based molecular communication \cite{Pierobon:2014iu}. Subsequently, we study the communication properties of the system, and its mutual information under different receiver configurations (series and parallel), as well as different numbers of receiver cells. Next we use the mutual information for different number of receiver cells in series or parallel and  compute the propagation speed by selecting a suitable threshold. We show that, in general, an increase in the number of receiving cells result in an increase in information propagation speed. 
\textcolor{black}{ The final aim of this work is to study the impact of an AP signal on the mutual information and propagation speed in the neighbourhood of the transmitting cell by comparing with the case when we have no AP signal. Furthermore we observe the effective range of AP signal. This suggests that it enables the plant to realize how far in the chain the information is transfered. By using this information a plant can make better decisions to coordinate among different cells and produce a physiological response (e.g. photosynthesis) to the external stimulus. }

We emphasize that throughout this paper, we consider the emission and propagation of {\em single AP signals}. While an AP can cause a neighbouring cell to also emit an AP, it is first essential to understand the behaviour of individual APs; we leave the analysis of cascading AP signals to future work.

The remainder of this paper is organized as follows. We describe the system model in Section \ref{system}. This includes the transmitter, action potential generation model and propagation medium model. Next we present the diffusion-only subsystem in Section \ref{diffusion}. This is followed by the modelling of the reaction-only subsystem in Section \ref{receiver}. The complete system model is presented in Section \ref{complete}. The expressions for mutual information and information propagation speed is derived in Section \ref{mutual}. Next we present the results of mutual information and information propagation speed for different number of receivers in series and parallel configurations in Section \ref{numerical}. \textcolor{black}{We also present the results for the impact of AP signal on the mutual information and information propagation of the system.} Finally Section \ref{conclusion} concludes the paper.

\begin{figure}
\begin{center}
\includegraphics[trim=0cm 0cm 0cm 0cm ,clip=true, width=0.9\columnwidth]{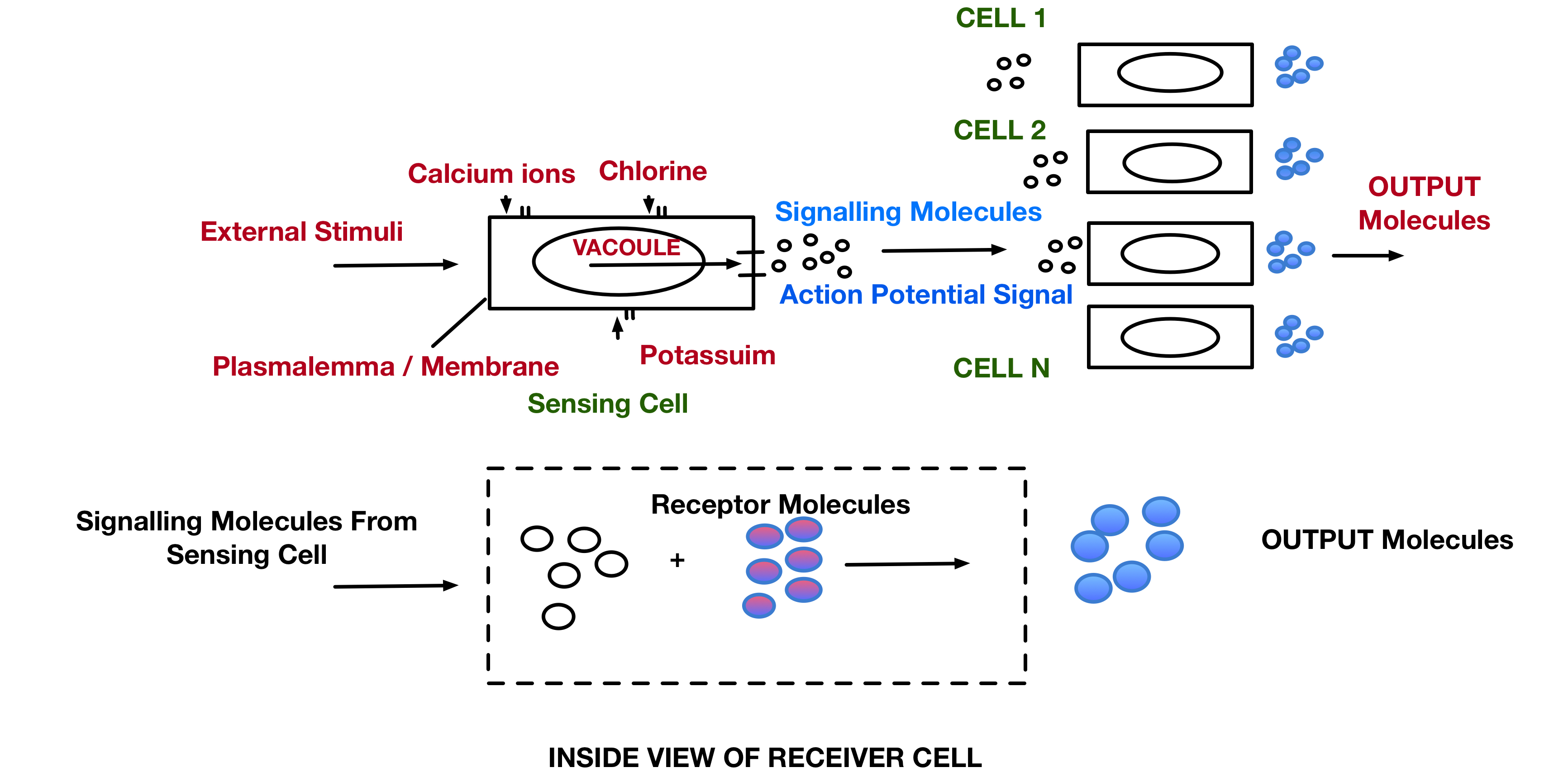}
\caption{System Model 1\textcolor{black}{: Parallel Configuration}}
\label{system parallel}
\end{center}
\end{figure}

\begin{figure}
\begin{center}
\includegraphics[trim=0cm 0cm 0cm 0cm ,clip=true, width=1\columnwidth]{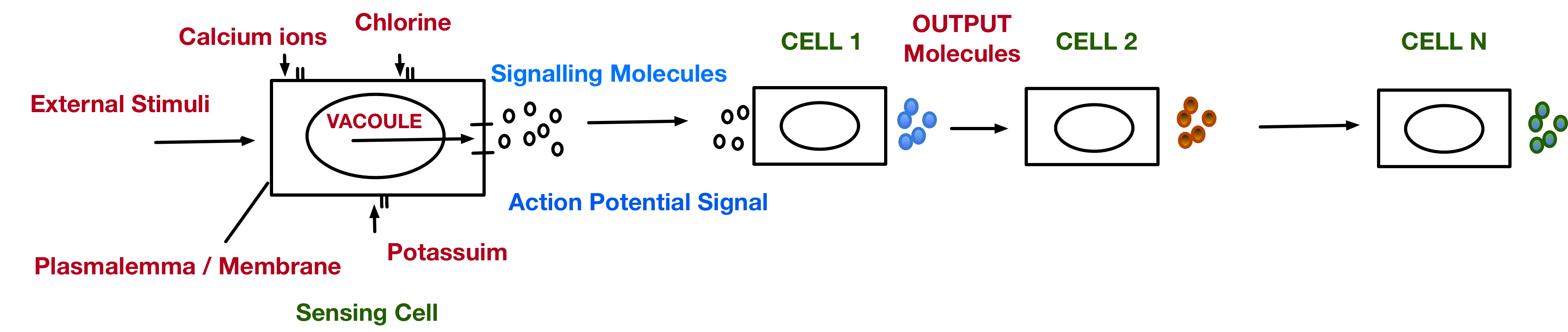}
\caption{System Model 2\textcolor{black}{: Series Configuration}}
\label{system series}
\end{center}
\end{figure}

\section{System Model}
\label{system}

In simple terms, the generation of an AP has three steps.
\begin{itemize}

\item Step 1: An AP signal is generated as a result of an external stimulus. The external stimulus causes the change in the parameters governing the AP generation. 

\item Step 2: \textcolor{black}{ As a result of the increase in membrane potential $E_m$, ion channels on the cell open in response to the AP signal, and the cell releases an increased number of molecules as compared to the case with no AP signal \cite{Awan18}.  This effect is similar to the neural action potential causing release of neurotransmitter molecules. } Thus, in the presence of an AP a high number of molecules are released as $E_m$ is high, but in the absence of AP signal few molecules are released as $E_m$ is small.

\item Step 3: \textcolor{black}{The molecules, released as a result of AP, propagate towards receiver cells via diffusion, where they are detected. (At the receiver, the detected molecules may cause the receiver cell to also emit an AP, though we will consider the cascade in a future paper.)}

\end{itemize}
From this simple description of our model, the key feature of the AP is the changing $E_m$; this leads to a significant increase in the number of emitted molecules compared with a situation where there is no AP.

As an overview of our techniques, the membrane potential in Steps 1 and 2 is modelled dynamically as a differential equation, and can be approximated by its steady-state value (after AP generation). In step 3, the propagation and reception can be modelled as a reaction-diffusion process, and we use discrete volume elements (voxels) to model the contents of cells as well as the medium of propagation.

\subsection{Sensing/Transmitter Cell}
\label{sec:transmitter}
\textcolor{black}{In a typical plant cell, the membrane of the cell has potential known as resting potential. A change in the resting potential can be induced by the external stimulus such as change in temperature, external electrical signal or damage by predation. This external stimulus results in the change of concentrations of ions in the outer cell membrane. The change in the different ion-concentrations such as  Ca$^{+2}$, Cl$^-$ and K$^+$ lead to the activation of potential-dependent ion-concentration channels.  As a result, there is an influx of ions into the cell which results in the development of the AP. Once this AP signal is generated in the system it triggers the release of increased signalling molecules from the sensing/transmitting cell. In this model we assume a diffusion based propagation of molecules from the transmitting cell to the receiver cells. Once these signal molecules reach the receiver they react with receptors to produce the output molecules. Based on this output signal we compute the mutual information between the input and output. }

\subsection{Action Potential Generation}
\textcolor{black}{We divide this section into two subsections. In the first subsection we will briefly describe the physics behind the generation of action potential in plants. This is followed by the mathematical model in the next section.} \\

\subsubsection{\textcolor{black}{Physics/Theory of Action Potential}}

The process of generation of AP signal in plant cell can be defined as follows. First, the external signal or stimulus causes a wave of positive charge on cell membrane which increases the resting potential of cell membrane. Next, the AP signal is generated when this resting potential crosses a certain threshold value, and as the potential rises, the ion channels open and ions such as Ca$^{+2}$ flow inside the cell. Finally, as the potential of the AP signal falls, the ions flow out of the cell, and as a result it returns to its resting potential. 

Normally the resting potential at the cell membrane is about -100-160 mV. The typical range of the AP signal is about 60-80 mV which is the increase in the resting potential. The cells in the plant are connected with each other, therefore any change in the resting potential of one cell will impact the neighboring cell. This can be explained in following way: The AP signal propagates to the cell boundary/membrane through the conducting vascular bundles which are believed to be AP propagation pathways in plants. On the cell membrane this AP signal triggers the release of increased number of signalling molecules which diffuse to the next cell. Therefore, this release of additional molecules is considered as the external input/stimulus to the next cell.
In the literature, some physical models are presented which consider the generation of AP under an external stimulus such as \cite{sukhov2009mathematical}. Some other work presents a model with an electrical processes developing on plasma-lemma of cell without stimulation \cite{gradmann1998electrocoupling}. \textcolor{black}{ There are some models which aim to present a theoretical and mathematical model of AP generation and propagation for plants \cite{sukhov2011simulation,sukhova2017mathematical}. }

\subsubsection{Mathematical Model}
Let $E_R$ represent the resting potential of the cell. The system is described by a set of equations which represent the change in membrane potential $E_m$ as a function of the changes in concentration of ions as \textcolor{black}{ given in \cite{sukhov2009mathematical,sukhov2011simulation}:}
\begin{equation}
\frac{dE_m}{dt} = \frac{1}{C} F \sum_{i}^{} z_i h_i , i  \in \textcolor{black}{({Ca^{+2}, Cl^-, K})}
\label{eq:1a}
\end{equation}
Where $F$ is Faraday's constant, $C$ is cell membrane capacitance, $z_i$ is ion $i$ charge and $h_i$ represents the ion flow in membrane as a result of change in concentrations. The expression for the ion flow $h_i$ is given as follows:
\begin{equation}
h_i = z  \mu P_m p_o \frac{\phi _i \eta_o - \phi_o \eta_i (\exp (-z \mu))}{1-\exp (-z \mu)} 
\end{equation}
where
\begin{equation}
 \mu = \frac{E_m F}{R_c T_c}
\end{equation}
To explain the terms in these equations: $z$ is ion charge; $\mu$ denotes the normalized resting potential; $R_c$ is the gas constant; $T_c$ is the temperature; $\phi_i$ (resp. $\eta_i$) is the probability that the ion is (resp. is not) linked to the channel on the inside; $\phi_o$ and $\eta_o$ are the corresponding terms for ions linked (or not) to the channel on the outside; $P_m$ represents the \textcolor{black}{maximum permeability of the cell}; and $p_o$ represents the ion-channel opening state probability as a result of change in concentrations of ions. For $k_1$ (opening)  and $k_2$ (closing) reaction rate constants we obtain $p_o$ as:
\begin{equation}
\frac{dp_o}{dt} = k_1 (1- p_o) - k_2 (p_o)
\end{equation}
\textcolor{black}{ Note that the channel $k_1$ (opening)  and $k_2$ (closing) reaction rate constants depend the on membrane potential crossing a specific threshold value, resulting in AP signal generation. We also note that these rate constants are exponentially dependent on the membrane potential similar to \cite{sukhova2017mathematical}. }

For simplicity we can replace differential Equation \eqref{eq:1a} for the membrane potential by following stationary equation as \textcolor{black}{ per the models given in \cite{sukhov2009mathematical,novikova2017mathematical}:}
\begin{equation}
E_m= \frac{g_k E_k + g_{cl} E_{cl} + g_{ca} E_{ca}}{g_k + g_{cl}+ g_{ca}}
\label{1:sa}
\end{equation}

\textcolor{black}{ Note that a proton pump is not included in this equation, as per the model in \cite{sukhov2011simulation}. The proton pump participates in the rest membrane potential generation, however, this equation is for the change in membrane potential as a result of flow of ions. }

\begin{figure}
\begin{center}
\includegraphics[trim=0cm 0cm 0cm 0cm ,clip=true, width=0.7\columnwidth]{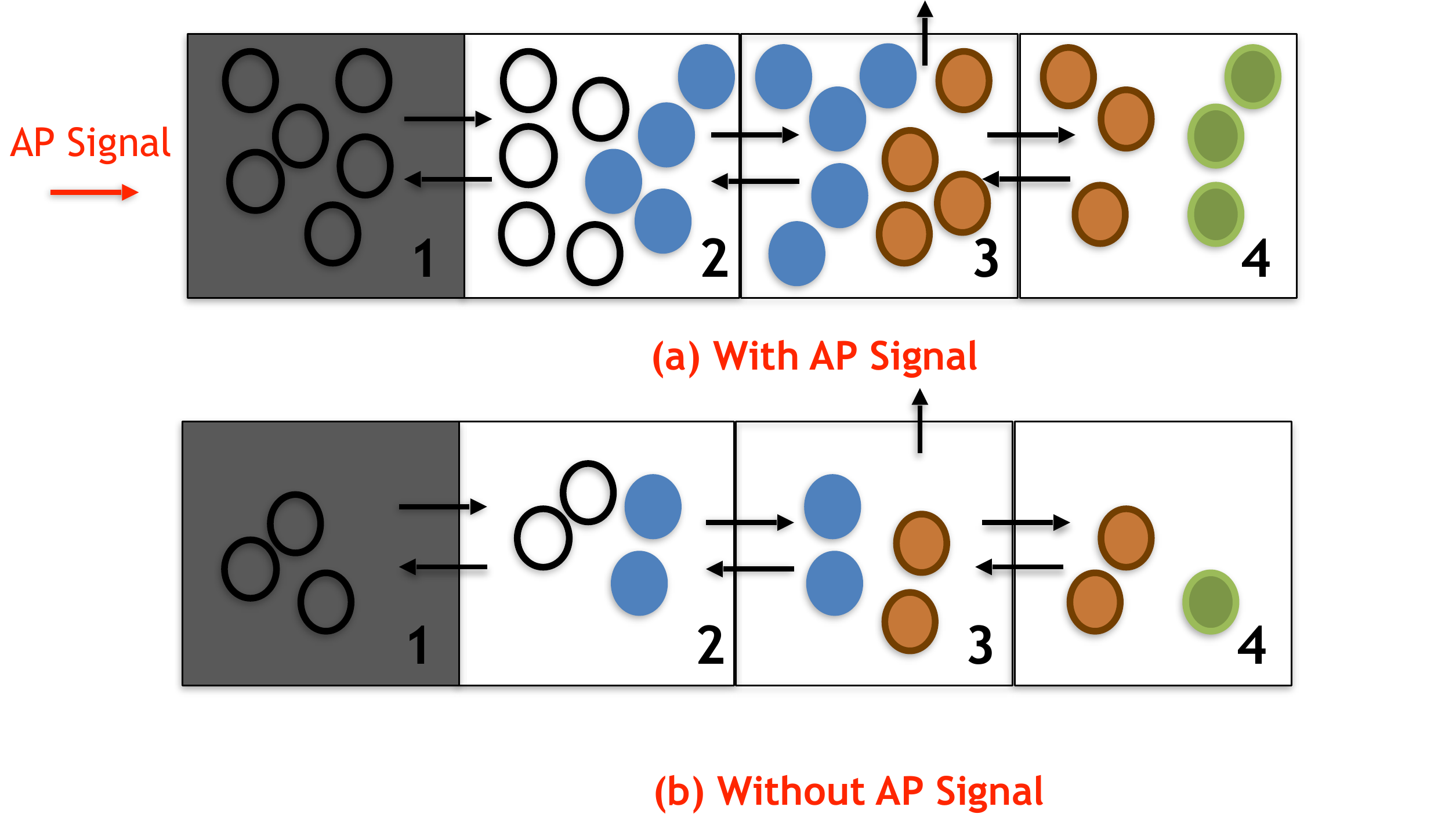}
\caption{Effect of AP Signal}
\label{AP effect}
\end{center}
\end{figure}

The term $g_i$ represents the electrical conductivity of the respective ion-channel and is given as:
 \begin{equation}
g_i = \frac{Fh_i}{E_R-E_i} 
\end{equation}
in which
$E_i$ represents the resting potential value for ion $i$, i.e., $E_k$ for potassium $K$ channel and similarly for other ions. The term $h_i$ represents the flux for the  ion channel $i$ and $E_R$ is resting potential. \textcolor{black}{The input of the system $U(t)$ i.e. the number of signalling molecules emitted by the transmitter/sensing cell is related to the membrane potential as}
\begin{equation}
U(t) \; \propto \; E_m
\label{1:ua}
\end{equation}
where $E_m$ represents the membrane potential. Note that this $U(t)$ acts as the system input in Section \ref{complete} and can be explained as $U(t) \in [h,l]$. This relation means that in the event of an AP signal generation the transmitter emits higher number of molecules as membrane potential is higher. Whereas, in case of no AP signal it releases fewer molecules as membrane potential is lower. To illustrate our approach in this section, as an example consider the case when we have a single transmitter cell and three receiver cells in series; see Figure \ref{AP effect}. We compare two cases here (a) diffusion of molecules from a transmitter to receivers in the presence of an AP signal and (b) diffusion of molecules in the absence of an AP signal. As shown in the Figure the presence of an AP signal increases the input signalling molecules and hence the output number of molecules.

\subsection{Propagation Mechanism}

In this section we explain the propagation mechanism for the signalling molecules from the sensing/transmitting cell to the receiver cell(s). The signalling molecules diffuse through the propagation medium and act as the input to the neighboring receiver cells.  We assume the medium of propagation is a three dimensional space of dimension $\ell_X \times \ell_Y \times \ell_Z$ where each dimension is an integral multiple of length $ \Delta$ i.e. there exist positive integers $M_x$, $M_y$ and $M_z$ such that $\ell_X = M_x \Delta$, $\ell_Y = M_y\Delta$ and $\ell_Z = M_z \Delta $. The medium is divided into $M_x \times M_y \times M_z$ cubic voxels where the volume of each voxel is $ \Delta^3$ and it represents a single cell.

\begin{figure}
\begin{center}
\includegraphics[trim=0cm 0cm 0cm 0cm ,clip=true, width=0.5\columnwidth]{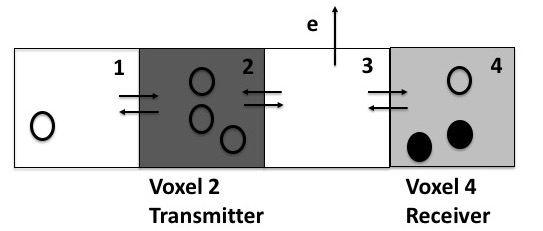}
\caption{Propagation Medium}
\label{1b}
\end{center}
\end{figure}

Figure \ref{1b} shows an example with $M_x$ = 4, $M_y$ =1 and $M_z$ = 1. \textcolor{black}{For the ease of presentation we describe this 1-dimensional example. An example for 2-dimensional case is shown later in this paper}. We assume that each voxel is given a unique index. The indices of the voxels are given in the top-right corner of the voxels in Figure \ref{1b}. We assume the transmitter and each receiver each occupy a single voxel. However, it is straightforward to generalize to the case where a transmitter or a receiver occupies multiple voxels. The transmitter and receiver are assumed to be located, respectively, at the voxels with indices $T$ and $R$. For example, in Figure \ref{1b}, Voxel 2 (dark grey) contains the transmitter and Voxel 4 (light grey) contains the receiver. Hence $T= 2$ and $R= 4$ for this example. The empty circles represent signaling or input molecules whereas the filled circles represent the output molecules. The different arrows in between voxels represents the  diffusion of molecules from one cell to another within the plant.

Diffusion is modelled by the molecules movement from one voxel to a neighbouring voxel. The arrows in Figure \ref{1b} show the directions of the movement of the molecules. We assume that the medium is homogeneous with the diffusion coefficient for the signalling molecule in the medium is $D$.  \textcolor{black}{The diffusion of molecules from a voxel to a neighbouring voxel takes place at a rate of $d = \frac{D}{\Delta^2}$. This means that within an infinitesimal time $\delta t$, the probability that a molecule diffuses to a neighbouring voxel is $d \delta t$. It is possible to model an inhomogeneous medium in this framework, see \cite{chou2015impact}, but we will not consider it here. }

\textcolor{black}{Furthermore, in this work we only consider intercellular communication for calculation of propagation speed. It is known that there are many types of plant cells which can be much larger, and hence, require intracellular communication to be considered for calculation of propagation speed. However, we leave this study for future work. }

\section{Reaction-Diffusion System}

In this paper we take the approach of dividing the system into two sub-systems, i.e., the diffusion only subsystem and the  reaction only system. We describe each subsystem in this section, and later describe how the systems are combined to form the complete communication model.

\subsection{Diffusion-Only System}
\label{diffusion}

The diffusion-only system considers the diffusions of signaling molecules from one voxel to another whereas the reaction only system considers the reactions taking place at the receiver voxel to produce the output molecules. This section explains how the diffusion of signaling molecules is modeled. 
 
Let $n_{L,i}$ denote the number of signaling molecules in the voxel $i$. In the absence of chemical reactions, the state of the system consists of only the number of signal molecules in each voxel. For the example in Figure \ref{1b} we have:
 \begin{equation}
n_L (t) = [n_{L,1}(t),n_{L,2}(t),n_{L,3}(t) ,n_{L,4}(t)]^T 
\label{1q:a}
\end{equation}
where the superscript $T$ denotes matrix transpose. Diffusion events can cause a change in the system state. For example the state of the system changes when a molecule diffuses from Voxel 1 to a neighboring Voxel 2 at a diffusion rate $dn_{L,1}$. This event causes $n_{L,1}$ to decrease by 1 and $n_{L,2}$ to increase by 1.  We can indicate this change by using jump vector $q_{d,1} (t) = [-1 ,1, 0, 0]^T$ where the subscript $d$ indicates that the jump vector originates from the diffusion of molecules and the subscript 1 is an index for this particular diffusion event. The state of system will be $n_L (t)+q_{d,1}$ after the occurrence of this diffusion event. We also specify the corresponding jump function $W_{d,1}(n_L (t))= dn_{L,1}$ which specifies the event rate. Similarly when the signaling molecule escapes out of the boundary of the medium, we use a jump vector dependent on rate of escape $e$. Let $J_d$ be the total number of diffusion events, then we have $J_d$ jump vectors $q_{d,j}$ and jump events $W_{d,j}(n_L (t))$ where $j = 1, ..., J_d$.

Combining the jump vectors and jump rate functions of all the diffusion and escape events we obtain a matrix $H$ for the medium in Eq \eqref{eqn:H}. For understanding the terms in $H$ matrix we present following explanation.  The $-d$ in row 1 column 1 means one ligand can diffuse out of voxel 1 hence negative sign. The $+d$ in row 1 column 2 means that one ligand molecules is diffusing in voxel 2 from voxel 1 hence positive sign. \textcolor{black}{Similarly in this example we allow the molecules to escape the medium through voxel 3. To model this we use $-e$.}
\begin{align}
H =  
\left[ \begin{array}{ccccc}
-d & d & 0 & 0  \\
d & -2d & d & 0  \\
0 & d & -2d-e & d   \\
0 & 0 & d & -d  \\
\end{array} \right] 
\label{eqn:H} 
\end{align}    
The diffusion events are stochastic and hence modeled by using a stochastic differential equation \cite{awan2017improving,gardiner2009stochastic} as follows:
\begin{align}
\dot{n}_L(t) & = \sum_{j = 1}^{J_d} q_{d,j}W_{d,j} (n_L(t)) + \sum_{j = 1}^{J_d} q_{d,j} \sqrt{W_{d,j}( n_L (t))} \gamma_j \nonumber \\
& + {\mathds 1}_T U(t)
\label{eqn:sde:do} 
\end{align}
\textcolor{black}{Note this is a form of chemical Langevin equation, in which $\gamma_j$ is continuous-time Gaussian white noise with unit power spectral density with $\gamma_{j1}$ independent of $\gamma_{j2}$ for $j_1$ $\neq$ $j_2$. This is similar to the model considered in \cite{awan2017improving}.}

There are three terms on the right-hand side of Eq.~\eqref{eqn:sde:do}, and we will discuss them one by one. The first term describes the deterministic dynamics. Since all the jump rates of all the diffusion events are linear, this term can be written as a product of a matrix $H$ and the state vector $n_L(t)$. 
\begin{align}
H n_L(t) & = \sum_{j = 1}^{J_d} q_{d,j}W_{d,j} (n_L(t))  
\end{align}
The second term of Eq.~\eqref{eqn:sde:do} describes the stochastic dynamics. An intuitive way to understand the first two terms in Eq.~\eqref{eqn:sde:do} is as follows. Over a finite time interval $\Delta t$, the number of times that the $j$-th type of jump occurs can be approximated by a Poisson random variable with mean $W_{d,j} (n_L(t)) \; \Delta t$. If $W_{d,j} (n_L(t)) \; \Delta t$ is large, then we know from probability theory that a Poisson random variable with mean $W_{d,j} (n_L(t)) \; \Delta t$ can be approximated by a Gaussian variable with both mean and variance given by $W_{d,j} (n_L(t)) \; \Delta t$. We can therefore approximate the number of times that the $j$-th type of jumps occurs by $W_{d,j} (n_L(t)) \; \Delta t + \sqrt{W_{d,j} (n_L(t)) \; \Delta t} \gamma_j$ where the first term is the mean number of jumps and the second term is the deviation from the mean; these two terms give rise to the first two terms in Eq.~\eqref{eqn:sde:do}. For a more detailed explanation, the reader can refer to \cite{Higham:2008dl}.

The third term models the transmitter. \textcolor{black}{ We model the transmitter by a function of time which specifies the emission rate of signalling molecules by the transmitter. ${\mathds 1}_T$ is a unit vector with 1 at the $T$-th element with the subscript $T$ being the index of the transmitter. We use $U(t)$ to denote the transmitter emission rate at time $t$. This means, in the time interval $[t,t+\delta t)$ the transmitter emits $U(t)\delta t$ signalling molecules. The input rate of signaling molecules i.e. $U(t)$ is proportional to $E_m$ as shown in Eq. \eqref{1:ua}. } Since the transmitter emits $U(t)\delta t$ molecules, we add this number of molecules to voxel $T$ (the index of the transmitter voxel) at time $t$.  The next step is to formulate the reaction only subsystem and to follow this up by combining both the subsystems to form a complete system.

\subsection{Reaction-Only Subsystem}
\label{receiver}

In this section we present a simple example of a receiver which consists of two linear chemical reactions described below.
We present the stochastic differential equation (SDE) governing the dynamics of a receiver, without considering diffusion. \textcolor{black}{ The reaction-only subsystem includes the reactions of incoming signaling molecules $L$ from the transmitter with the receiver to produce output molecules $X$.} The count of these output molecules over time is the output signal of the system. The signaling molecules go through a number of reactions at the receiver voxel. Each reaction will be described by its chemical formula (on the left-hand side), and jump vector and jump rate (on the right-hand side).
\begin{align}
L &  \rightarrow X 		& \left[ \begin{array}{cc} -1 & 1 \end{array} \right]^T&, k_+ n_{L,R}  \label{cr:rc1}  \\
X & \rightarrow L		& \left[ \begin{array}{cc} 1 & -1 \end{array} \right]^T&, k_- n_X       \label{cr:rc2}
\end{align}
\textcolor{black}{In the above equations, description, $n_{L,R}$ and $n_X$ denote, respectively, the number of signalling molecules in the receiver voxel and the output molecules. The symbols $k_+$ and $k_-$ denote the reaction rate constants. } Note that the scalar term $n_{L,R}$ differs from the vector $n_{L}$ which refers to the number of signalling molecules in all the voxels as shown in Equation \eqref{1q:a}.

In reaction \eqref{cr:rc1} the signaling molecules react at rate ${k}_{+} n_{L,R}$ to produce output molecules. The change in number of signaling and output molecules is indicated by jump vectors. Since the signaling molecules are consumed, the first term of the jump vector is $-1$. Likewise, the output molecules are produced, so the second term of the jump vector is $+1$. Similarly we can understand the jump vector entries for  reaction \eqref{cr:rc2}. We can model the reaction  only system using stochastic differential equations for different receiver reactions. Note that the input  is $ n_{L,R} $  i.e the number of signaling molecules. The output of this subsystem is  the number of output molecules $ n_{X} $. The state vector and SDE for the reaction only system is given as:
\begin{align}
 \tilde{n}_R(t) 
 & =  \left[ \begin{array}{c|c}
 n_{L,R}(t) & n_X(t)  
\end{array} \right]^T 
\end{align} 
\begin{align}
\dot{\tilde{n}}_R(t) & = R_v \tilde{n}_R(t) + \sum_{j = J_d+1}^{J_d + J_r} q_{r,j} \sqrt{W_{r,j}(\langle \tilde{n}_R(t) \rangle)} \gamma_j 
\label{eqn:sde:ro11} 
\end{align}

Like the modeling of the diffusion-only subsytem we use jump vectors $q_{r,j}$ and jump rates $W_{r,j}$ to model the reactions. $\gamma_j$ represents the continuous time white noise. The reactions are indexed from $J_d+1$ to $ J_d + J_r$ where  $J_d$ is  for the diffusion only module and $J_r$ represents the reactions in the receiver. \textcolor{black} { We define the matrix $R_v$ as a  2$\times$2 block matrix and its entries depend on the reactions of signaling molecules in the receiver. The matrix $R
_v$ for the above reactions is given in Table \ref{table:1s}.  In first reaction the signaling molecule $L$ is consumed so $R_{11}$ $=$ $-k_+$. Whereas in second reaction the output molecules revert to signaling molecules so $R_{12}$ $=$  $k_-$. In the same way we obtain $R_{21}= k_+ $ and $R_{22}= -k_- $.}  In the next section we combine the diffusion only and reaction only modules to obtain a  diffusion-reaction combined system.

\begin{table}[]
\centering
\caption{$R_v$ Matrix for Receiver Reactions}
\begin{tabular}{|c|c|}
\hline
\multicolumn{1}{|c|}{Receiver }	&	\multicolumn{1}{|c|}{$R_v$ Matrix}	\\
\hline
Receiver Reactions &    $ \begin{bmatrix}  -k_{+} & k_{-} \\ k_{+} & -k_{-} \end{bmatrix}$
\\ \hline
\end{tabular}
\label{table:1s}
\end{table}

\subsection{Reaction-Diffusion Combined System}
\label{complete}

In this section, we will combine the SDE models for the diffusion-only subsystem and the reaction-only subsystem to form the complete system model. We developed two subsystems separately so that the behavior of the combined system can be expressed in terms of the interconnection of these two subsystems. Note that the number of signaling molecules  $n_{L,R}(t)$ appears in the state vectors $n_L(t)$ and $\tilde{n}_R(t)$ of both diffusion only and receiver only modules respectively.  Therefore, the interconnection between the diffusion-only subsystem and the reaction-only subsystem is the number of signaling molecules in the receiver voxel, which is common to both of them.

We will use the example in Figure \ref{1b} to explain how the diffusion-only subsystem and the reaction only system can be combined together. We consider the dynamics of the diffusion-only subsystem for  the example, when the receiver voxel has the index $R = 4$. The evolution of the number of signaling molecules in the receiver voxel $n_{L,R}(t)$ is given by the $R$-th (i.e. fourth) row of Eq.~\eqref{eqn:sde:do}, which is: 
\begin{align}
& \dot{n}_{L,R}(t) = d n_{L,3}(t) - d n_{L,R}(t) \nonumber \\
& + \underbrace{\sum_{j = 1}^{J_d} [q_{d,j}]_R \sqrt{W_{d,j}(n_L(t))} \gamma_j}_{\xi_d(t)}
\label{eqn:sde:ro:r2}  
\end{align}
where $[q_{d,j}]_R$ denote the $R$-th element of the vector $q_{d,j}$. \textcolor{black}{Note that there is no $e$ in this equation as the model described in Figure \ref{1b} does not allow molecules to escape from receiver voxel $R = 4$ as seen in last row of the $H$ matrix. }The dynamics of the number of signaling molecules in the receiver voxel due to the reactions in the receiver is given by the first element of Eq.~\eqref{eqn:sde:ro11}, i.e.: 
 \begin{align}
& \dot{n}_{L,R}(t) = R_{11} n_{L,R}(t) + R_{12} n_X(t)  \nonumber \\
&+ \underbrace{\sum_{j = J_d+1}^{J_d+J_r} [q_{r,j}]_1 \sqrt{W_{r,j}(\tilde{n}_R(t))} \gamma_j}_{\xi_r(t)} 
\label{eqn:sde:ro:r1a} 
\end{align}
where $[q_{r,j}]_1$ denotes the first element of the vector $q_{r,j}$.
For the complete system the dynamics of $n_{L,R}(t)$ is obtained by combining Eqs. \eqref{eqn:sde:ro:r2} and \eqref{eqn:sde:ro:r1a} as follows: 
\begin{align}
\dot{n}_{L,R}(t) = &  d n_{L,3}(t) - d n_{L,R}(t) + R_{11} n_{L,R}(t)  + R_{12} n_X(t) \nonumber \\
& + \xi_{total}(t) 
\label{eqn:sde:nlr:ex} 
\end{align}
where $\xi_{total}(t) = \xi_d(t) + \xi_r(t)$.  We now describe the complete model. Let $n(t)$ be the state of the complete system and it is given by: 
\begin{align}
n(t) = 
 & \left[ \begin{array}{c|c}
 n_{L}(t)^T & n_X(t)  
\end{array} \right]^T
\label{eqn:state} 
\end{align}
We will also need to modify the jump vectors from the diffusion-only subsystem and the reaction-only subsystem, to obtain the jump vectors for the complete model.  We use $q_j$ and $W_j(n(t))$ to denote the jump vectors and jump rates of the combined model. The SDE for the complete system is:
\begin{align}
\dot{n}(t) & = A n(t) + \sum_{i = 1}^{J} q_j \sqrt{W_j(n(t))} \gamma_j + {\mathds 1}_T U(t) 
\label{eqn:mas11}
\end{align} 
where $J = J_d+J_r$, and the matrix $A$ is defined by $A n(t) = \sum_{i = 1}^{J} q_j W_j(n(t))$. The matrix $A$ has the block structure: 
\begin{align}
A = 
 & \left[ \begin{array}{c|c}
H + {\mathds 1}_R^T {\mathds 1}_R R_{11}  &   {\mathds 1}_R R_{12}  \\ \hline 
R_{21}  {\mathds 1}_R^T & R_{22} 
\end{array} \right]
\label{eqn:A} 
\end{align}
where $H$ comes from the diffusion only subsystem and we can interpret this matrix as the infinitesimal generator of a Markov chain which describes the diffusion of molecules. Similarly $R_{11}$, $R_{12}$ etc come from the reaction only subsystem. The vector ${\mathds 1}_R$ is a unit vector with an 1 at the $R$-th position; in particular, note that  ${\mathds 1}_R n_L(t) = n_{L,R}(t)$ which is the number of signalling molecules in the receiver voxel. Note that, the coupling between the diffusion-only subsystem and the output module, as exemplified by Eq. \eqref{eqn:sde:nlr:ex}, takes place at the $R$-th row of $A$. 

We now explain how the jump vectors for the complete system are formed. \textcolor{black} {Let $m_d$ and  $m_r$  denote the dimension of the vector $n_L(t)$ and $n_X(t)^T$.} Note that $m_d$ is in fact the number of voxels. The dimension of the jump vectors $q_j$ in the complete system is $m_d+1$. Given a jump vector $q_{d,j}$ ($j = 1,...,J_d$) from the diffusion only subsystem with dimension $m_d$, we append a zero to $q_{d,j}$ to obtain $q_j$. The jump vectors $q_{r,j}$ ($j = J_d+1,...,J_d+J_r$) from the output module have dimension $m_r+1$. To obtain $q_j$ from $q_{r,j}$, we do the following: (1) take the first element of $q_{r,j}$ and put it in the $R$-th element of $q_j$; (2) take the last element of $q_{r,j}$ and put it in the last element of $q_j$. Note that jump rates are unchanged when combining the subsystems.

Next we obtain the Laplace transform of the number of signaling molecules in the receiver for the combined system which will then lead to the number of output molecules of the system.
\begin{align}
  N(s)  = 
 \underbrace{   (sI - A)^{-1} {\mathds 1}_T }_{\Psi(s)}   U(s)
\label{eqn:mas11}
\end{align}
\begin{align}\mbox{where,  }
\Psi(s) = \frac{G_{XL} (s) H_{RT} (s)}{1 - (R_{11} + G_{LL} (s) )H_{RR} (s)} 
\end{align}
\begin{align}
& H_{RT}(s) = {\mathds 1}_R^T (sI - H)^{-1} {\mathds 1}_T , \nonumber \\
& G_{LL}(s) = R_{12}(sI - R_{22})^{-1} R_{21} \nonumber \\
& G_{XL}(s) = {\mathds 1}_X^T(sI - R_{22})^{-1} R_{21} ,\nonumber \\
& H_{RR}(s) = {\mathds 1}_R^T (sI - H)^{-1} {\mathds 1}_R
\end{align}
Where $H_{RT}(s)$ is the Laplace transform of the probability that a molecule emitted by  transmitter $T$ at time $t=0$ is in the  receiver $R$ at time $t$. The function $H_{RR}(s)$ is the Laplace transform of the probability that a signaling molecule present in the receiver voxel R at time $t= 0 $ is present in the receiver voxel R at time $t$.  The transfer functions $G_{XL}(s)$ and $G_{LL}(s)$ are generated by the receiver only module. $G_{XL}(s)$ is the Laplace transform of the probability that an output molecule $X $ at time $t$ is produced by a signaling molecule $L$ at time $t=0$. Similarly $G_{LL}(s)$ is the Laplace transform of the probability that a signaling molecule $L$ in the receiver at time $t$ is produced by a signaling molecule $L$ in the receiver at time $t=0$ through the reactions with the receiver circuit. Therefore the transfer function $\Psi(s)$  takes into account the consumption of signaling molecules, the interaction between output molecules and the signaling molecules, as well as the possibility that a signaling molecule may leave or return in the receiver. 

\section{Mutual Information and Information Propagation}
\label{mutual}

We divide this section into two subsections. First we compute mutual information, and subsequently derive the speed of information propagation.

\subsection{Mutual Information}
The input and output signals for the complete system are, respectively, the production rate $U(t)$ of the signalling molecules in the transmitter voxel and the number of output molecules $n_X(t)$ in the receiver voxel. In this section, we will derive an expression for the mutual information between the input $U(t)$ and output $n_X(t)$. We begin by stating a result in \cite{tostevin2010mutual}, for two Gaussian distribution random processes $a(t)$ and $b(t)$, their mutual information $I(a,b)$ is given by: 
\begin{align}
I(a,b) &= \frac{-1}{4\pi} \int_{-\infty}^{\infty} \log \left( 1 - \frac{|\Phi_{ab}(\omega)|^2}{\Phi_{aa}(\omega) \Phi_{bb}(\omega)}  \right) d\omega
\label{eqn:MI0}
\end{align} 
where $\Phi_{aa}(\omega)$ (resp. $\Phi_{bb}(\omega)$) is the power spectral density of $a(t)$ ($b(t)$), and $\Phi_{ab}(\omega)$ is the cross spectral density of $a(t)$ and $b(t)$. In order to apply the above results to the communication link given in Eq.~\eqref{eqn:mas11}, we need a result from \cite{warren2006exact} on the power spectral density of systems consisting only of chemical reactions with linear reaction rates. Following from \cite{warren2006exact} if all the jump rates $W_j(n(t))$ in Eq. \eqref{eqn:mas11} are linear in $n(t)$, \textcolor{black}{ then the power spectral density of $n(t)$ is obtained by using the following SDE:}
\begin{align}
\dot{n}(t) & = A n(t) + \sum_{i = 1}^{J} q_j \sqrt{W_j(\langle n(\infty) \rangle)} \gamma_j + {\mathds 1}_T U(t) 
\label{eqn:complete2}
\end{align} 
where $\langle n(t) \rangle$ denotes the mean of $n(t)$ and is the solution to the following ordinary differential equation:
\begin{align}
\dot{\langle n(t) \rangle} & = A \langle n(t) \rangle+ {\mathds 1}_T c 
\label{eqn:ode_ninfinity}
\end{align} 
where $c$, which was defined before, is the mean of input $U(t)$.  As a result, the dynamics of the complete system in Eq.~\eqref{eqn:complete2} are described by a set of  linear SDE with $U(t)$ as the input and $n_X(t)$ (which is the last element of the state vector $n(t)$) as the output. \textcolor{black} {The summation term on the right-hand side of Eq.~\eqref{eqn:complete2} is used to account for the noise in the system due to diffusion and reactions.} The input $U(t)$ has the form $U(t) = c + w(t)$ where $c$ (mean of input) is dependent on $E_m$ and $w(t)$ is a zero-mean Gaussian random process. The noise in the output $n_X(t)$ is caused by the Gaussian white noise variables $\gamma_j$ in Eq.~\eqref{eqn:complete2}. Therefore, Eq.~\eqref{eqn:complete2} models a continuous-time linear time-invariant (LTI) stochastic system subject to Gaussian input and Gaussian noise.

\textcolor{black}{The power spectral density $\Phi_{X}(\omega)$ of the signal $n_X(t)$ can be obtained from standard results on the output response of a LTI system to a stationary input \cite{Papoulis} and is given by: }
\begin{align}
\Phi_{{X}}(\omega) & =  |\Psi(\omega) |^2 \Phi_e(\omega) + \Phi_{\eta}(\omega)
\label{2331a} 
\end{align}
where $\Phi_e(\omega)$ is the power spectral density of $U(t)$ and $|\Psi(\omega)|^2$ is the channel gain with $\Psi(\omega) = \Psi(s)|_{s = i\omega}$ defined by:
\begin{align}
 \textcolor{black}{\Psi(s) = {\mathds 1}_X   (sI - A)^{-1} {\mathds 1}_T}   
\label{21a}
\end{align}
Note that Eq.~\eqref{21a} can be obtained from Eq.~\eqref{eqn:complete2} after taking the mean and applying the Laplace transform as explained in previous section. The term $\Phi_{\eta}(\omega)$ denotes the stationary noise spectrum and is given by: 
\begin{align}
\Phi_{\eta}(\omega) & =   \sum_{j = 1}^{J_d + J_r} | {\mathds 1}_X (i \omega I - A)^{-1} q_j |^2 W_j(\langle n_{}(\infty) \rangle) 
\label{eqn:spec:noise2} 
\end{align} 
where $n_{}(t)$ denotes the state of the complete system in Eq. \eqref{eqn:state} and $\langle n_{}(\infty) \rangle$ is the mean state of system at time $\infty$ due to constant input $c$. Similarly, by using standard results on the LTI system, the cross spectral density $\Psi_{xe}(\omega)$ has the following property:
\begin{align}
|\Psi_{xu}(\omega)|^2 &= |\Psi(\omega) |^2 \Phi_e(\omega)^2 
\label{eqn:csd} 
\end{align}

By substituting Eq.~\eqref{2331a} and Eq.~\eqref{eqn:csd} in the mutual information expression in Eq.~\eqref{eqn:MI0}, we arrive at the mutual information $I(n_{X},U)$ between $U(t)$ and $n_{X}(t)$ is:
\begin{align}
I(n_{X},U) = \frac{1}{2} \int \log \left( 1+\frac{ | \Psi(\omega) |^2}{\Phi_{\eta}(\omega)} \Phi_e(\omega) \right) d\omega
\label{eqn:mi1}
\end{align}
The maximum mutual information of the communication link can be determined by applying the water-filling solution to Eq. \eqref{eqn:mi1} subject to a power constraint on input $U(t)$ \cite{gallager1968information}. 

\subsection{Information Propagation Speed}
\label{infops}
In this section we discuss how we use of the mutual information to obtain the relationship between information propagation speed and number of cells in the chain. First we obtain the mutual information for different number of receiver cells in series or parallel. The next step is to choose a suitable threshold value so that we can calculate the time difference at which the mutual information curve for each case crosses the threshold value. Next we use the following equation for calculating the information propagation speed $V$ (cells/sec):
\begin{equation}
  V=  \frac{1}{\mathbf{E} [\Delta t_{i,i+1}]}
  \label{infop}
\end{equation}
Note that in this equation we are using unit distance, hence no distance term.
where $\Delta t_{i,i+1}$ represents the time difference at which the mutual information for each case (i.e. increasing receivers) crosses the threshold value. \textcolor{black}{ $\mathbf{E}$ denotes the expectation operator.} This technique is used to compute the propagation speed for an increasing number of receiver cells in the chain in series. We present the results for this approach in numerical examples section.

\section{Series vs. Parallel Receiver Configurations}

As shown in Figures \ref{system parallel} and  \ref{system series} we consider a single transmitter and a number of receiver cells in either configurations.
The system model described above is for the case when we have one sensing/transmitter cell and one receiver cell as shown in Figure \ref{1b}. The $H$ matrix shown in Equation \eqref{eqn:H} is for this case as well. Here we analyze to the series configuration in Figure \ref{1b}, and also analyze a configuration of receivers in parallel with  each other.

\subsection{Series Case}
Now we consider the receiver configuration for the case when we have receiver cells in series. For this case the configuration of voxels in the propagation medium is shown in Figure \ref{1b}. 
 
Keeping the earlier mentioned explanation of propagation medium in mind we have assumed 9 voxels in the system such that $M_x = 4$, $M_y =1$ and $M_z = 1$. The indices of the voxels are given in the top-right corner of the voxels in Figure \ref{1b}. We assume the transmitter occupies the voxel 1 whereas we have three receivers in series in voxels 2, 3 and 4 respectively. Hence $T = 1$ and $R = 2,3,4 $ for this example. The empty circles represent signaling or input molecules whereas the colored circles represent the different output molecules. Note that for this configuration we have 3 different types of output molecules. This is because the signaling molecules $L$ released from the main sensing cell react with the first receiver cell of the chain to produce output molecules $X_{[1]} $ (blue molecules in the figure). The output molecules $X_{[1]}$ of the first receiver cell then react with the the second receiver cell (in chain) to produce $X_{[2]}$ (brown molecules in the figure). Similarly, the $X_{[2]}$ of the second receiver cell react with the third receiver cell to produce $X_{[3]}$ green molecules in the next receiver cell. The reactions are:
\begin{align}
\cee{
L &<=>C[k_+][k_-] X_{[1]}\nonumber \\
X_{[1]}  &->C[k_*] X_{[2]}\ \nonumber\\
X_{[2]} &->C[k_**] X_{[3]}\ 
 \label{eq:numd}  }
\end{align}

In Figure \ref{1b} the different arrows between voxels represent the  diffusion of molecules from one cell to another within the plant.
Recall the $H$ matrix from \eqref{eqn:H}, corresponding to Figure \ref{1b}.

For this case we use the same approach as described above for a single receiver cell. However in this case we will have a series of differential equations representing the change in the number of output molecules at the end of each receiver cell. These equations have the same form as that for a single receiver cell given in Eq. \eqref{eqn:mas11}. For each of these differential equation the $A$ matrix will be different. We can compute the mutual information and information propagation speed for these cases in similar way as for one receiver cell. The expression for mutual information will be of the same form as mutual information expression for the single receiver cell in Eq. \eqref{eqn:mi1}.

\begin{figure}
\begin{center}
\includegraphics[trim=0cm 0cm 0cm 0cm ,clip=true, width=0.5\columnwidth]{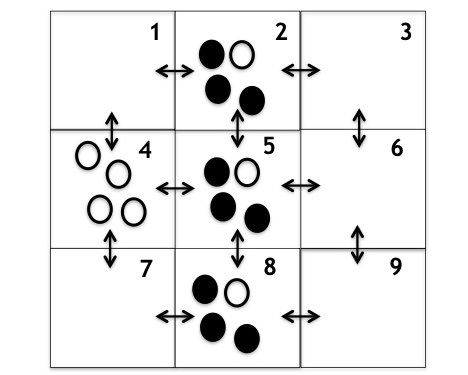}
\caption{Propagation Medium Parallel System}
\label{1c}
\end{center}
\end{figure}

\subsection{Parallel Case}

First we describe the case where we consider a number of receivers in parallel. For this case the configuration of voxels in the propagation medium is shown in Figure \ref{1c}. Keeping the earlier mentioned explanation of propagation medium in mind we have assumed 9 voxels in the system such that $M_x$ = 3, $M_y$ =3 and $M_z$ = 1. The indices of the voxels are given in the top-right corner of the voxels in Figure \ref{1c}. We assume the transmitter occupies the voxel 4 whereas we have three receivers in parallel in voxels 2, 5 and 8 respectively. Hence $T = 4$ and $R = 2,5,8 $ for this example. The empty circles represent signaling or input molecules whereas the filled circles represent the output molecules. The different arrows in between voxels represents the  diffusion of molecules from one cell to another within the plant. For this scheme the $H$  matrix is will be 9 $\times$ 9 and is given as follows:

\begin{gather}
\scalebox{0.85}{$
\begin{aligned}
H =  
\left[ \begin{array}{ccccccccc}
-2d & d & 0  & d & 0 & 0 & 0 & 0 & 0\\
d & -3d & d  & 0 & d & 0 & 0 & 0 & 0\\
0 & d & -2d & 0 & 0 & d & 0 & 0 & 0\\
d & 0 & 0  & -3d & d & 0 & d & 0 & 0\\
0 & d & 0  & d & -4d & d & 0 & d & 0\\
0 & 0 & d  & 0 & d & -3d & 0 & 0 & d\\
0 & 0 & 0  & d & 0 & 0 & -2d & d & 0\\
0 & 0 & 0  & 0 & d & 0 & d & -3d & d\\
0 & 0 & 0  & 0 & 0 & d & 0 & d & -2d
\end{array} \right] 
\label{eqn:H2} 
\end{aligned}$} 
\end{gather}

\begin{table*}[t]
\centering
\caption{Parameters and their default values.}
\begin{tabular}{|c|c|}
\hline
\multicolumn{1}{|c|}{Symbols}	&	\multicolumn{1}{|c|}{Notation and Value}	
\\ \hline
$E_R$ &     Resting Potential of cell membrane  = -150-170 Milli volts.
\\ \hline
$F$ &    Faraday's constant = $9.65 x 10^4 C/mol$ 
\\ \hline
$C$ &     Membrane capacity = $10^{-6} F cm^-2$  
\\ \hline
$P_m$  &     Permeability per unit area  = $10^{-6}$ M cm $s^{-1}$
\\ \hline
$\gamma$ &  ratio of association-disassociation rate constants = $9.9 x 10^-5 M$ 
\\ \hline
 $\phi_{i}$ & Probability ion link to channel inside $c_{in}$ / ($c_{in}$ + $\gamma$)
\\ \hline
 $\phi_{o}$ & Probability ion link to channel outside $c_{out}$ / ($c_{out}$ + $\gamma$)
 \\ \hline
 $\eta_i$ & Probability ion not linked to channel inside = 1- $\phi_{i}$ 
\\ \hline
 $\eta_o$ & Probability ion not linked to channel outside = 1- $\phi_{o}$
\\ \hline
$c_{in}$ and $c_{out}$ &  1.28	and 1.15 respectively.
\\ \hline
$z$   & ion charge e.g. for calcium = +2  
\\ \hline
$p_o$ & ion channel open-state probability. 
\\ \hline
\end{tabular}
\label{table:1}
\end{table*}
  
Similarly for the reactions at the receiver we have same reaction at all three parallel receiver cell locations.
\begin{align}
\cee{
L &<=>C[k_+][k_-] X 
}
\end{align}
Note that as a result of three parallel receivers in the system the number of output molecules will change as compared to single receiver case. Our aim is to compute the  mutual information and information propagation speed. We use the same approach for multiple number of receivers as we used for the single receiver. The approach explained in the previous section is general and therefore can be readily applied to any number of receivers in the system.

\begin{figure}
\begin{center}
\includegraphics[trim=0cm 0cm 0cm 0cm ,clip=true, width=1\columnwidth]{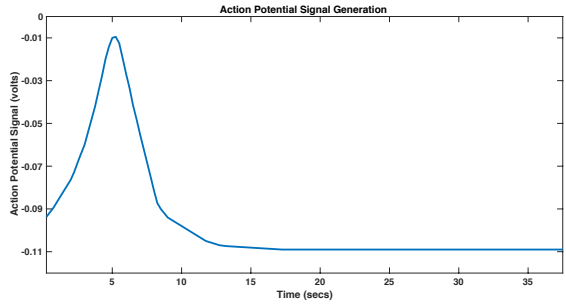}
\caption{\textcolor{black}{Action Potential Generation}}
\label{Result 2AP1}
\end{center}
\end{figure}

\section{Numerical Results}
\label{numerical}
\begin{figure}
\begin{center}
\includegraphics[trim=0cm 0cm 0cm 0cm ,clip=true, width=1\columnwidth]{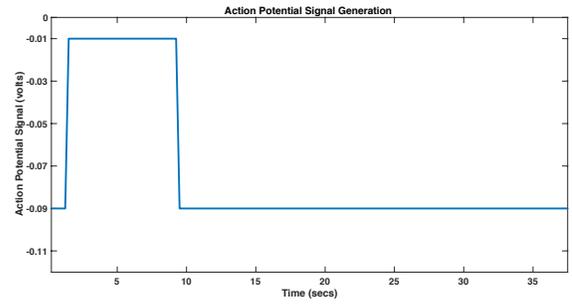}
\caption{\textcolor{black}{Action Potential Generation-Steady State Approximation}}
\label{Result 2AP}
\end{center}
\end{figure}

In this section we discuss the numerical results related to this work. The parameter selection for the generation of AP signal in given in \textcolor{black}{Table \ref{table:1}.}  Using these parameter values, the results in \textcolor{black}{Figures \ref{Result 2AP1}} and  \ref{Result 2AP} show that the magnitude of AP signal generated is about 60-80 millivolts. The result in Figure \ref{Result 2AP1} shows the actual AP signal whereas the result in Figure \ref{Result 2AP} is a steady state approximation of the AP signal. Once the AP signal is generated it triggers the release of an increased number of signalling molecules from the cell membrane. A typical AP signal will release a constant number of molecules through it, i.e., approximately 20-25 per second. For the result in Figure \ref{Result 2AP} the external stimulus has a start point at $t=1$ and end point at $t=10$ with peak at $t=5$. The AP signal in response to the external stimulus response in the same pattern with peak at $t=5$ and eventually returning to resting potential value at time $t=10$.

Next we describe the parameters for the propagation medium of this system. In this work we assume a voxel size of ($\frac{1}{3}$$\mu$m)$^{3}$ (i.e. $\Delta = \frac{1}{3}$ $\mu$m), creating an array of $4\times 1 \times 1$ voxels for the series configuration. For parallel configuration of receivers we assume the an array of  $3\times 3 \times 1$ voxels.  The transmitter and each receiver receiver occupy one voxel each as mentioned in system model. We assume the diffusion coefficient $D$ of the medium is $1$ $\mu$m$^2$s$^{-1}$. The deterministic emission rate $c$ is dependent on the AP signal. 

\begin{figure}
\begin{center}
\includegraphics[trim=0cm 0cm 0cm 0cm ,clip=true, width=1\columnwidth]{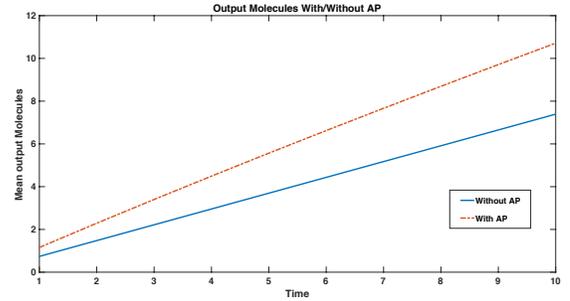}
\caption{\textcolor{black}{Number of Output Molecules in presence of AP}}
\label{Result 6}
\end{center}
\end{figure}

 \textcolor{black}{The literature shows that cell (voxel) size of plants vary across a wide range for different species from very few $\mu$m$^{3}$ upto ten of $\mu$m$^{3}$. In this work we have chosen this value as an example, however, the generalization of this model with different cell sizes is straightforward. Since this paper considers only linear reactions, the choice of voxel size does not significantly affect the simulation results. For further discussion on the effect of voxel size, see \cite{Chou:rdmex_arxiv} which is an extended version of  \cite{Chou:rdmex_nc}. Similarly, the diffusion co-efficients in different plant species vary across a wide range.  For the model presented in the paper, it can be varied according to the plant species. This will, however, not impact the general result patterns shown in this paper.}

\begin{figure}
\begin{center}
\includegraphics[trim=0cm 0cm 0cm 0cm ,clip=true, width=1\columnwidth]{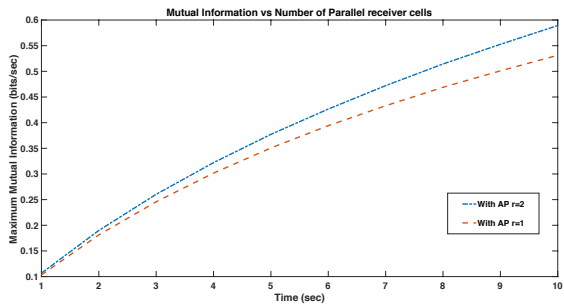}
\caption{\textcolor{black}{Mutual Information for Parallel Case 2 Receivers}}
\label{Result 7}
\end{center}
\end{figure}

The aim is to study the relation between the input and output number of molecules of the complete system for different receiver configurations i.e. receivers in parallel and receiver in series. Note that for receiver reactions we select the reaction rate constants $k_+$ = $k_-$ =  1. \textcolor{black}{ Note that the results in this section i.e. Figures 8-13 correspond to the scheme of each different models presented in Figures  \ref{system parallel} and \ref{system series}.}

Next we present the result for the mean number of output molecules in the presence of AP signal. We obtain this result for the system with single sensing cell and single receiving cell. The system uses random diffusion and ligand receptor binding reactions at the receiver end. \textcolor{black}{Figure \ref{Result 6}} shows that the number of output molecules \textcolor{black}{tends} to increase in the presence of the AP signal. We note that this figure corresponds to the AP signal shown in Figure \ref{Result 2AP} where the signal is present from time $t=1-10$.

\begin{figure}
\begin{center}
\includegraphics[trim=0cm 0cm 0cm 0cm ,clip=true, width=1\columnwidth]{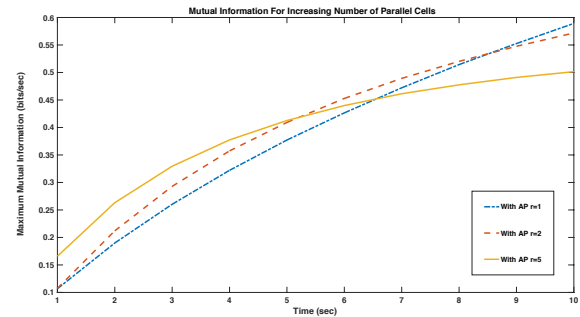}
\caption{\textcolor{black}{Mutual Information for Parallel Case 5 Receivers}}
\label{Result 2}
\end{center}
\end{figure}

The next step is to use the number of signaling molecules as the input from the sensing cell to the receiving cells. First we consider the case with a single sensing/transmitting cell and two receiver cells in parallel. The mutual information between the input signaling molecules and combined number of output molecules for this case is shown in \textcolor{black}{Figure \ref{Result 7}.} In the next step we increase the number of cells at receiver to study the impact of increasing number of cells on the mutual information. Keeping the same parallel configuration of the receiver we now consider a single sensing/transmitting cell and five receiver cells in parallel. The result for the mutual information between the input and output molecules for this case is shown in \textcolor{black}{ Figure \ref{Result 2}.}

\textcolor{black}{ We realize that increasing the number of parallel cells in the receiver side we observe an increase in the maximum mutual information initially, but as the time progresses the mutual information with higher number of cells in parallel tends to decrease as compared to a smaller number of cells. To explain this intuitively we present following example: When the number of parallel receiver cells increase, the number of signalling molecules available to each receiver cell tends to decrease. This is due to the increased noise and random flow of  molecules to all the receivers. This leads to a slightly lesser number of output molecules and hence reduction in mutual information. } \textcolor{black}{This result suggests an important insight into the working of plant cells. It suggests that in the case of parallel receiver the impact of AP signal will only be effective for a certain number of cells. This information can help the plant to send further AP signals to the remaining cells in order to execute a coordinated response (such as photosynthesis) to the external stimulus.}

\begin{figure}
\begin{center}
\includegraphics[trim=0cm 0cm 0cm 0cm ,clip=true, width=0.9\columnwidth]{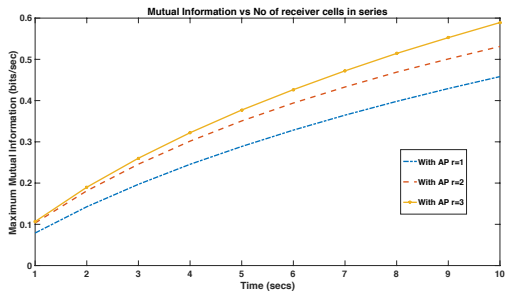}
\caption{Mutual Information For Series Case 3 Receivers}
\label{Result 5}
\end{center}
\end{figure}

The next results are for the series configuration of receiver cells with a single sensing/transmitting cell. In this case we use three receiver cells in series forming a chain of cells. For this case the mutual information results in \textcolor{black}{ Figure \ref{Result 5}} show that the mutual information tends to increase with the increase in the number of cells in series. We further study the cases when we increase the number of receiver cells in series upto $5$ and we obtain similar results. The results are not included due to lack of space.

In this case we observe that the maximum mutual information tends to increase with the increase in the number of cells. This result is different to the parallel case and mutual information tends to increase for high number of receivers. An intuitive explanation for this in the series setting of the receivers, the AP signal is directly impacting only one receiver cell so the number of signalling molecules are not being divided into multiple receivers. For a single receiver the number of signalling molecules are enough to keep producing higher number of output molecules in each cell till the end of the chain.

\begin{figure}
\begin{center}
\includegraphics[trim=0cm 0cm 0cm 0cm ,clip=true, width=0.8\columnwidth]{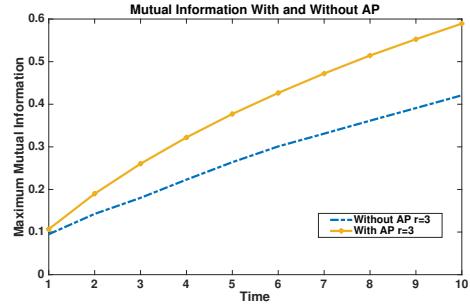}
\caption{\textcolor{black}{Mutual Information (bits/sec) With and Without AP}}
\label{Ap-m}
\end{center}
\end{figure}

\begin{figure}
\begin{center}
\includegraphics[trim=0cm 0cm 0cm 0cm ,clip=true, width=0.8\columnwidth]{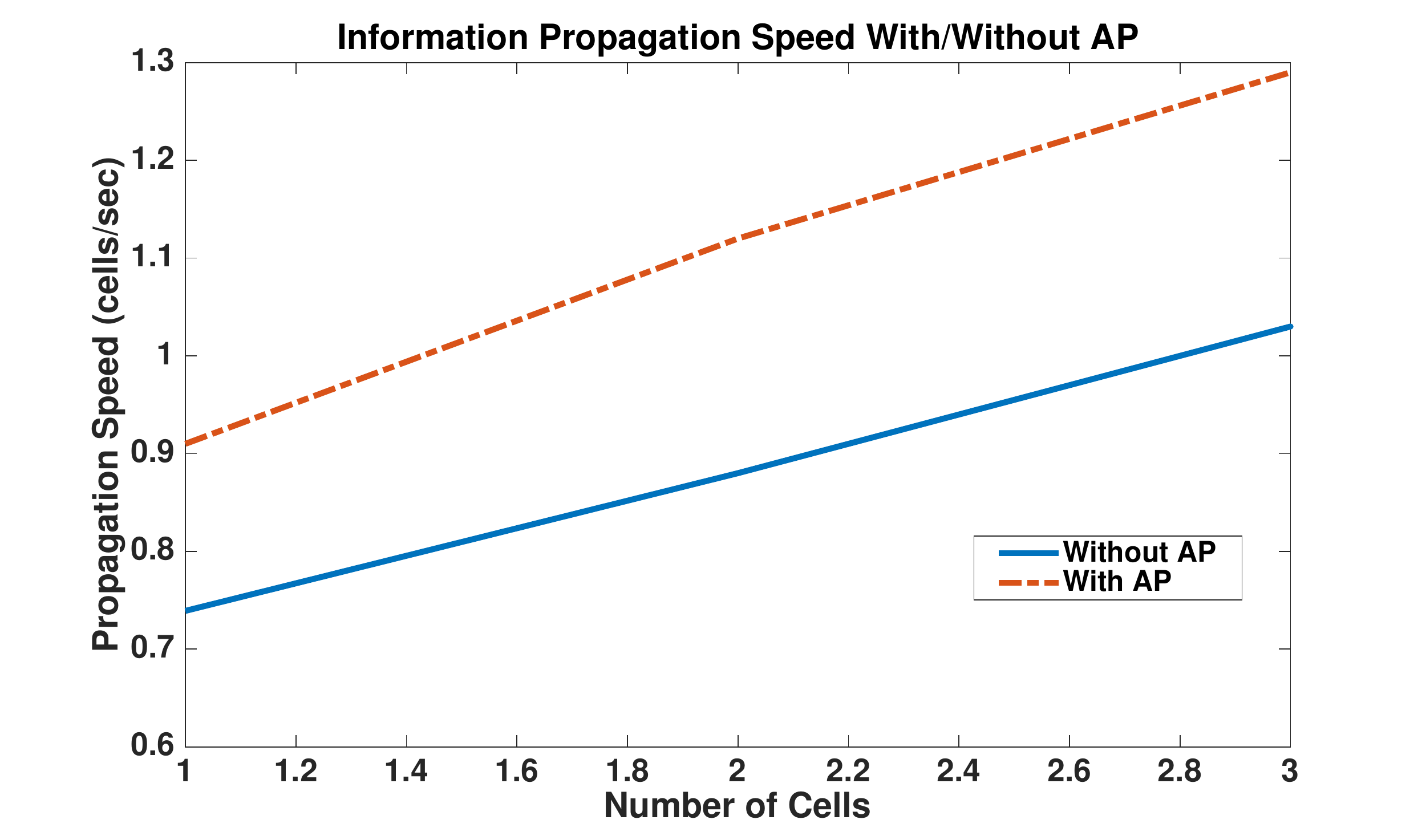}
\caption{\textcolor{black}{Information Propagation Speed With and Without AP-Series Case}}
\label{Ap-ip}
\end{center}
\end{figure}

\textcolor{black}{ Next we present another interesting result where we show the impact of AP signal on the mutual information. In Figure \ref{Ap-m} we show that the mutual information increases in the presence of AP signal for the system with single transmitter and three receiver cells in series. This result holds for both the series and parallel configuration of the receiver. However due to limited space we only show one result as an example. } \textcolor{black}{Next we study the impact of AP and increasing number of cells in the chain on the information propagation rate. By using the mutual information and selecting a threshold value we show that the information propagation speed increases in the presence of an AP signal as shown in Figure \ref{Ap-ip}. Note that this result for the case when we have single transmitter and three receivers in series. This result holds for both the series and parallel configuration of the receiver. However due to limited space we only show one result as an example.}

We also study the variation in the information propagation speed for increasing  number of receiver cells in both  series and parallel settings. \textcolor{black}{ The mechanisms of these effects, and the measurement technique, are presented in Section \ref{infops} and Equation \eqref{infop}. An intuitive explanation for the increase in information propagation speed with the increase in the number of series cells is that the number of molecules adding at each cell are generated by each cell independently.}

\begin{figure}
\begin{center}
\includegraphics[trim=0cm 0cm 0cm 0cm ,clip=true, width=0.85\columnwidth]{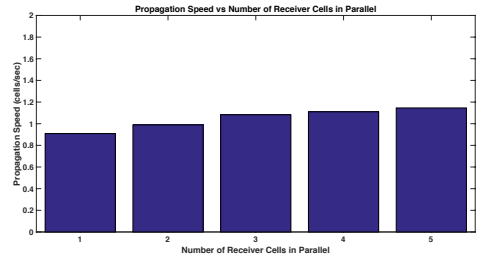}
\caption{Propagation Speed vs No. of cells parallel}
\label{Result 3}
\end{center}
\end{figure}

\begin{figure}
\begin{center}
\includegraphics[trim=0cm 0cm 0cm 0cm ,clip=true, width=0.85\columnwidth]{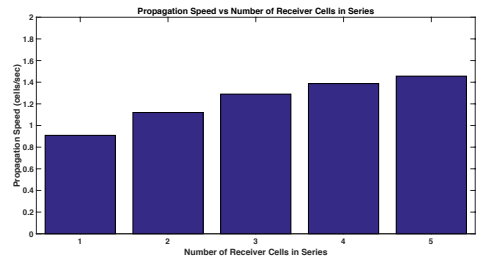}
\caption{Propagation Speed vs No. of cells series}
\label{Result 9}
\end{center}
\end{figure}

The result in Figure \ref{Result 3} shows that the information propagation speed tends to increase with the increase in the number of parallel cells. Similarly the result in Figure \ref{Result 9} show that shows that the information propagation speed tends to increase with the increase in the number of series cells. Hence we can conclude that information propagation speed in general, tends to increase with increasing number of receiver cells.

\section{Conclusion}
\label{conclusion}

 In this paper we present a general model for the generation of Action Potential signals in higher organisms such as plants. We note that the AP signal triggers the release of an increased number of signalling molecules from the transmitting cell towards the receiving cell(s). When these molecules arrive at the receiver cell they react to produce the output molecules which form the output signal of the system. In this work for two different receiver configurations i.e., series and parallel we computed the mutual information between the input signal from transmitting cell and output signal of the receiver cell(s). As expected, we observe that the mean number of output molecules increase in the presence of AP signal. we also conclude that the mutual information and the information propagation speed tends to increase in the presence of AP signal.

\textcolor{black}{ The results in this work are significant from a biological perspective. Our results present the first quantification of information propagation in plant action potential signals. We observe that the mutual information and the information propagation speed also increase in general, with the increase in the number of receiver cells.  The work of this paper also suggests that by knowing the effective range of propagation of AP signal, a plant (transmitter cell) can determine how far in the chain the information is transferred. This helps the plant to resend the information to remaining cells, and therefore efficiently coordinate a physiological response (such as photosynthesis) to external stimulus.}

Future work will include a study of the propagation mechanism of the AP signal as an electrical signal as compared to electro-chemical signal, and a study of the cascade as APs induce further APs in neighbouring cells.

\appendices

\section*{Acknowledgment}
Funding for this research was provided by the Defense Advanced Research Projects Agency (DARPA) RadioBio program.


\ifCLASSOPTIONcaptionsoff
  \newpage
\fi

\bibliographystyle{ieeetr}
\bibliography{nano2017,book,nano2018a,newstuff}

\begin{thebibliography}{10}

\bibitem{bose1914}
J.~C. Bose, ``{An automatic method for the investigation of velocity of
  transmission of excitation in {\em Mimosa}},'' {\em Phil. Trans. B},
  vol.~204, pp.~63--97, 1914.

\bibitem{pickard1973}
B.~G. Pickard, ``Action potentials in higher plants,'' {\em Botanical Review},
  vol.~39, no.~2, pp.~172--201, 1973.

\bibitem{sukhov2009mathematical}
V.~Sukhov and V.~Vodeneev, ``A mathematical model of action potential in cells
  of vascular plants,'' {\em Journal of Membrane Biology}, vol.~232, no.~1-3,
  p.~59, 2009.

\bibitem{kress2009action}
G.~J. Kress and S.~Mennerick, ``Action potential initiation and propagation:
  upstream influences on neurotransmission,'' {\em Neuroscience}, vol.~158,
  no.~1, pp.~211--222, 2009.

\bibitem{fromm2007electrical}
J.~Fromm and S.~Lautner, ``Electrical signals and their physiological
  significance in plants,'' {\em Plant, cell \& environment}, vol.~30, no.~3,
  pp.~249--257, 2007.

\bibitem{sukhov2011simulation}
V.~Sukhov, V.~Nerush, L.~Orlova, and V.~Vodeneev, ``Simulation of action
  potential propagation in plants,'' {\em Journal of Theoretical Biology},
  vol.~291, pp.~47--55, 2011.

\bibitem{fromm1995biochemical}
J.~Fromm, M.~Hajirezaei, and I.~Wilke, ``The biochemical response of electrical
  signaling in the reproductive system of hibiscus plants,'' {\em Plant
  physiology}, vol.~109, no.~2, pp.~375--384, 1995.

\bibitem{sukhov2016electrical}
V.~Sukhov, ``Electrical signals as mechanism of photosynthesis regulation in
  plants,'' {\em Photosynthesis Research}, vol.~130, no.~1-3, pp.~373--387,
  2016.

\bibitem{szechynska2017electrical}
M.~Szechy{\'n}ska-Hebda, M.~Lewandowska, and S.~Karpi{\'n}ski, ``Electrical
  signaling, photosynthesis and systemic acquired acclimation,'' {\em Frontiers
  in physiology}, vol.~8, p.~684, 2017.

\bibitem{surova2016variation}
L.~Surova, O.~Sherstneva, V.~Vodeneev, L.~Katicheva, M.~Semina, and V.~Sukhov,
  ``Variation potential-induced photosynthetic and respiratory changes increase
  atp content in pea leaves,'' {\em Journal of plant physiology}, vol.~202,
  pp.~57--64, 2016.

\bibitem{sukhov2015variation}
V.~Sukhov, L.~Surova, O.~Sherstneva, A.~Bushueva, and V.~Vodeneev, ``Variation
  potential induces decreased psi damage and increased psii damage under high
  external temperatures in pea,'' {\em Functional Plant Biology}, vol.~42,
  no.~8, pp.~727--736, 2015.

\bibitem{sukhov2017high}
V.~Sukhov, V.~Gaspirovich, S.~Mysyagin, and V.~Vodeneev, ``High-temperature
  tolerance of photosynthesis can be linked to local electrical responses in
  leaves of pea,'' {\em Frontiers in physiology}, vol.~8, p.~763, 2017.

\bibitem{lautner2014involvement}
S.~Lautner, M.~Stummer, R.~Matyssek, J.~Fromm, and T.~E. Grams, ``Involvement
  of respiratory processes in the transient knockout of net co2 uptake in m
  imosa pudica upon heat stimulation,'' {\em Plant, cell \& environment},
  vol.~37, no.~1, pp.~254--260, 2014.

\bibitem{pena1995signals}
H.~Pe{\~n}a-Cort{\'e}s, J.~Fisahn, and L.~Willmitzer, ``Signals involved in
  wound-induced proteinase inhibitor ii gene expression in tomato and potato
  plants,'' {\em Proceedings of the National Academy of Sciences}, vol.~92,
  no.~10, pp.~4106--4113, 1995.

\bibitem{hlavavckova2006electrical}
V.~Hlav{\'a}{\v{c}}kov{\'a}, P.~Krch{\v{n}}{\'a}k, J.~Nau{\v{s}}, O.~Nov{\'a}k,
  M.~{\v{S}}pundov{\'a}, and M.~Strnad, ``Electrical and chemical signals
  involved in short-term systemic photosynthetic responses of tobacco plants to
  local burning,'' {\em Planta}, vol.~225, no.~1, p.~235, 2006.

\bibitem{sukhova2018influence}
E.~Sukhova, M.~Mudrilov, V.~Vodeneev, and V.~Sukhov, ``Influence of the
  variation potential on photosynthetic flows of light energy and electrons in
  pea,'' {\em Photosynthesis research}, vol.~136, no.~2, pp.~215--228, 2018.

\bibitem{sherstneva2015participation}
O.~Sherstneva, V.~Vodeneev, L.~Katicheva, L.~Surova, and V.~Sukhov,
  ``Participation of intracellular and extracellular ph changes in
  photosynthetic response development induced by variation potential in pumpkin
  seedlings,'' {\em Biochemistry (Moscow)}, vol.~80, no.~6, pp.~776--784, 2015.

\bibitem{beilby1984current}
M.~Beilby, ``Current-voltage characteristics of the proton pump atchara
  plasmalemma: I. ph dependence,'' {\em The Journal of Membrane Biology},
  vol.~81, no.~2, pp.~113--125, 1984.

\bibitem{beilby2016re}
M.~J. Beilby and S.~Al~Khazaaly, ``Re-modeling chara action potential: I. from
  thiel model of ca2+ transient to action potential form.,'' {\em AIMS
  Biophysics}, vol.~3, no.~3, pp.~431--449, 2016.

\bibitem{mummert1991action}
H.~Mummert and D.~Gradmann, ``Action potentials inacetabularia: Measurement and
  simulation of voltage-gated fluxes,'' {\em The Journal of membrane biology},
  vol.~124, no.~3, pp.~265--273, 1991.

\bibitem{sukhova2017mathematical}
E.~Sukhova, E.~Akinchits, and V.~Sukhov, ``Mathematical models of electrical
  activity in plants,'' {\em The Journal of membrane biology}, vol.~250, no.~5,
  pp.~407--423, 2017.

\bibitem{felle2007systemic}
H.~H. Felle and M.~R. Zimmermann, ``Systemic signalling in barley through
  action potentials,'' {\em Planta}, vol.~226, no.~1, p.~203, 2007.

\bibitem{novikova2017mathematical}
E.~Novikova, V.~Vodeneev, and V.~Sukhov, ``Mathematical model of action
  potential in higher plants with account for the involvement of vacuole in the
  electrical signal generation,'' {\em Biochemistry (Moscow), Supplement Series
  A: Membrane and Cell Biology}, vol.~11, no.~2, pp.~151--167, 2017.

\bibitem{evans2017chemical}
M.~J. Evans and R.~J. Morris, ``Chemical agents transported by xylem mass flow
  propagate variation potentials,'' {\em The Plant Journal}, vol.~91, no.~6,
  pp.~1029--1037, 2017.

\bibitem{vodeneev2018parameters}
V.~Vodeneev, M.~Mudrilov, E.~Akinchits, I.~Balalaeva, and V.~Sukhov,
  ``Parameters of electrical signals and photosynthetic responses induced by
  them in pea seedlings depend on the nature of stimulus,'' {\em Functional
  Plant Biology}, vol.~45, no.~2, pp.~160--170, 2018.

\bibitem{nakano2013-book}
T.~Nakano, A.~W. Eckford, and T.~Haraguchi, {\em Molecular communication}.
\newblock Cambridge University Press, 2013.

\bibitem{farsad2016comprehensive}
N.~Farsad, H.~B. Yilmaz, A.~Eckford, C.-B. Chae, and W.~Guo, ``A comprehensive
  survey of recent advancements in molecular communication,'' {\em IEEE
  Communications Surveys and Tutorials}, vol.~18, no.~3, pp.~1887--1919, 2016.

\bibitem{Akyildiz:2008vt}
I.~Akyildiz, F.~Brunetti, and C.~Bl{\'a}zquez, ``{Nanonetworks: A new
  communication paradigm},'' {\em Computer Networks}, vol.~52, pp.~2260--2279,
  2008.

\bibitem{Hiyama:2010jf}
S.~Hiyama and Y.~Moritani, ``{Molecular communication: Harnessing biochemical
  materials to engineer biomimetic communication systems},'' {\em Nano
  Communication Networks}, vol.~1, pp.~20--30, May 2010.

\bibitem{Nakano:2014fq}
T.~Nakano, T.~Suda, Y.~Okaie, M.~J. Moore, and A.~V. Vasilakos, ``{Molecular
  Communication Among Biological Nanomachines: A Layered Architecture and
  Research Issues},'' {\em NanoBioscience, IEEE Transactions on}, vol.~13,
  no.~3, pp.~169--197, 2014.

\bibitem{gilroy2016ros}
S.~Gilroy, M.~Bia{\l}asek, N.~Suzuki, M.~G{\'o}recka, A.~R. Devireddy,
  S.~Karpi{\'n}ski, and R.~Mittler, ``Ros, calcium, and electric signals: key
  mediators of rapid systemic signaling in plants,'' {\em Plant Physiology},
  vol.~171, no.~3, pp.~1606--1615, 2016.

\bibitem{Pierobon:2010kz}
M.~Pierobon and I.~Akyildiz, ``{A physical end-to-end model for molecular
  communication in nanonetworks},'' {\em IEEE JOURNAL ON SELECTED AREAS IN
  COMMUNICATIONS}, vol.~28, no.~4, pp.~602--611, 2010.

\bibitem{farsad2011simple}
N.~Farsad, A.~W. Eckford, S.~Hiyama, and Y.~Moritani, ``A simple mathematical
  model for information rate of active transport molecular communication,'' in
  {\em Computer Communications Workshops (INFOCOM WKSHPS), 2011 IEEE Conference
  on}, pp.~473--478, IEEE, 2011.

\bibitem{Chou:2014jca}
C.~T. Chou, ``{Molecular communication networks with general molecular circuit
  receivers},'' in {\em ACM The First Annual International Conference on
  Nanoscale Computing and Communication}, (New York, New York, USA), pp.~1--9,
  ACM Press, 2014.

\bibitem{awan2016generalized}
H.~Awan and C.~T. Chou, ``Generalized solution for the demodulation of reaction
  shift keying signals in molecular communication networks,'' {\em IEEE
  Transactions on Communications}, 2016.

\bibitem{awan2016demodulation}
H.~Awan and C.~T. Chou, ``Demodulation of reaction shift keying signals in
  molecular communication network with protein kinase receiver circuit,'' in
  {\em Wireless Communications and Networking Conference (WCNC), 2016 IEEE},
  IEEE, 2016.

\bibitem{Awan:2016:RER:2967446.2967455}
H.~Awan, ``Reducing the effect of reaction rate constants on the performance of
  molecular communication networks,'' in {\em Proceedings of the 3rd ACM
  International Conference on Nanoscale Computing and Communication},
  NANOCOM'16, (New York, NY, USA), pp.~8:1--8:6, ACM, 2016.

\bibitem{Awan:2015:IRM:2800795.2800798}
H.~Awan and C.~T. Chou, ``Impact of receiver molecular circuits on the
  performance of reaction shift keying,'' in {\em Proceedings of the Second
  Annual International Conference on Nanoscale Computing and Communication},
  NANOCOM' 15, (New York, NY, USA), pp.~2:1--2:6, ACM, 2015.

\bibitem{riaz2018using}
M.~U. Riaz, H.~Awan, and C.~T. Chou, ``Using spatial partitioning to reduce
  receiver signal variance in diffusion-based molecular communication,'' in
  {\em Proceedings of the 5th ACM International Conference on Nanoscale
  Computing and Communication}, p.~12, ACM, 2018.

\bibitem{Pierobon:2014iu}
M.~Pierobon, ``{A Molecular Communication System Model Based on Biological
  Circuits},'' in {\em ACM The First Annual International Conference on
  Nanoscale Computing and Communication}, (New York, New York, USA), pp.~1--8,
  ACM Press, 2014.

\bibitem{Awan18}
H.~Awan, R.~S. Adve, N.~Wallbridge, C.~Plummer, and A.~W.~Eckford,
  ``Characterizing information propagation in plants,'' {\em Accepted for
  Publication in IEEE Globecom Conference AbuDhabi, UAE}, Dec 2018.

\bibitem{gradmann1998electrocoupling}
D.~Gradmann and J.~Hoffstadt, ``Electrocoupling of ion transporters in plants:
  interaction with internal ion concentrations,'' {\em The Journal of membrane
  biology}, vol.~166, no.~1, pp.~51--59, 1998.

\bibitem{chou2015impact}
C.~T. Chou, ``Impact of receiver reaction mechanisms on the performance of
  molecular communication networks,'' {\em Nanotechnology, IEEE Transactions
  on}, vol.~14, no.~2, pp.~304--317, 2015.

\bibitem{awan2017improving}
H.~Awan and C.~T. Chou, ``Improving the capacity of molecular communication
  using enzymatic reaction cycles,'' {\em IEEE transactions on nanobioscience},
  2017.

\bibitem{gardiner2009stochastic}
C.~Gardiner, {\em Stochastic methods}.
\newblock Springer Berlin, 2009.

\bibitem{Higham:2008dl}
D.~J. Higham, ``{Modeling and Simulating Chemical Reactions},'' {\em SIAM
  Review}, vol.~50, no.~2, p.~347, 2008.

\bibitem{tostevin2010mutual}
F.~Tostevin and P.~R. Ten~Wolde, ``Mutual information in time-varying
  biochemical systems,'' {\em Physical Review E}, vol.~81, no.~6, p.~061917,
  2010.

\bibitem{warren2006exact}
P.~B. Warren, S.~T{\u{a}}nase-Nicola, and P.~R. ten Wolde, ``Exact results for
  noise power spectra in linear biochemical reaction networks,'' {\em The
  Journal of chemical physics}, vol.~125, no.~14, p.~144904, 2006.

\bibitem{Papoulis}
A.~Papoulis and S.~U. Pillai, {\em Probability, Random Variables and Stochastic
  Processes}.
\newblock McGraw Hill, 2002.

\bibitem{gallager1968information}
R.~G. Gallager, {\em Information theory and reliable communication}, vol.~2.
\newblock Springer, 1968.

\bibitem{Chou:rdmex_arxiv}
C.~Chou, ``Extended master equation models for molecular communication
  networks,'' Tech. Rep. arXiv:1204.4253, arXiv, 2012.

\bibitem{Chou:rdmex_nc}
C.~T. Chou, ``Noise properties of linear molecular communication networks,''
  {\em Nano Communication Networks}, vol.~4, pp.~87--97, 2013.

\end{thebibliography}
\begin{IEEEbiography}[{\includegraphics[width=1in,height=1.25in,clip,keepaspectratio]{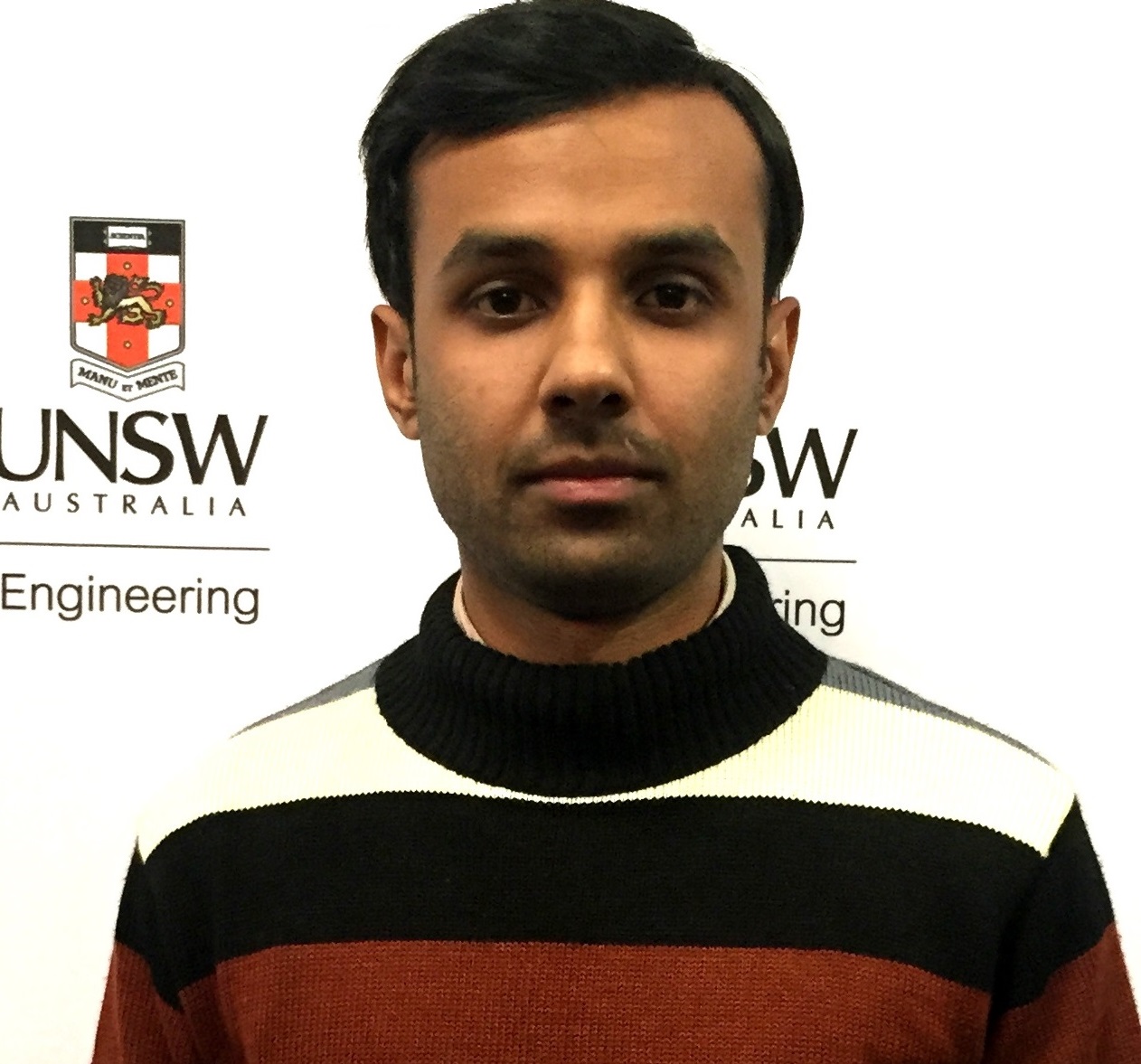}}] {Hamdan Awan} (S’14) received his BSc and MSc degrees in Electrical engineering from University of Engineering and Technology, Taxila, Pakistan in 2011 and 2013, respectively. He has completed his Ph.D. degree from the School of Computer Science and Engineering, The University of New South Wales (UNSW), Sydney, Australia in August 2017. He continued as a Post doctoral research fellow at UNSW from September 2017 to November 2017. Since December 2017 he has joined the Department of Electrical Engineering and Computer Science at York University, Toronto, Canada where he is funded by DARPA'S Radio Bio program. His current research interests are molecular communication, nano-networks and information theory aspects of biological communication.
\end{IEEEbiography}
\vskip -2\baselineskip plus -1fil
\begin{IEEEbiography}[{\includegraphics[width=1in,height=1.25in,clip,keepaspectratio]{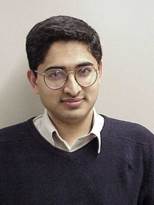}}] 
{Raviraj S. Adve} (S'88, M '97, SM’06, F’17) was born in Bombay, India. He received his B. Tech. in Electrical Engineering from IIT, Bombay, in 1990 and his Ph.D. from Syracuse University in 1996, His thesis received the Syracuse University Outstanding Dissertation Award. Between 1997 and August 2000, he worked for Research Associates for Defense Conversion Inc. on contract with the Air Force Research Laboratory at Rome, NY. He joined the faculty at the University of Toronto in August 2000 where he is currently a Professor. Dr. Adve’s research interests include molecular communications, analysis and design techniques for cooperative and heterogeneous networks, energy harvesting networks and in signal processing techniques for radar and sonar systems. He received the 2009 Fred Nathanson Young Radar Engineer of the Year award. Dr. Adve is a Fellow of the IEEE.
\end{IEEEbiography}
\vskip -2\baselineskip plus -1fil
\begin{IEEEbiography}[{\includegraphics[width=1in,height=1.25in,clip,keepaspectratio]{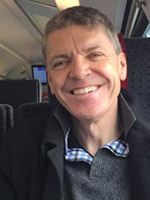}}] {Nigel Wallbridge} is an internationally renowned serial entrepreneur with extensive experience founding and growing companies in both North America and Europe. A chartered engineer, he holds a doctoral degree in medical engineering from the University of Leeds in the UK and also an MBA from INSEAD, France.
Apart from his current work at Vivent, he is also currently an Executive Board Member at TeleRail Networks Ltd and is on the board of a number of other technology businesses. He previously held senior managerial positions in the telecommunications industry, including CEO of Interoute, and President of Cable and Wireless Americas. He also lectured in medical engineering at the University of Leeds.
\end{IEEEbiography}
\vskip 0pt plus -1fil
\begin{IEEEbiography}[{\includegraphics[width=1in,height=1.25in,clip,keepaspectratio]{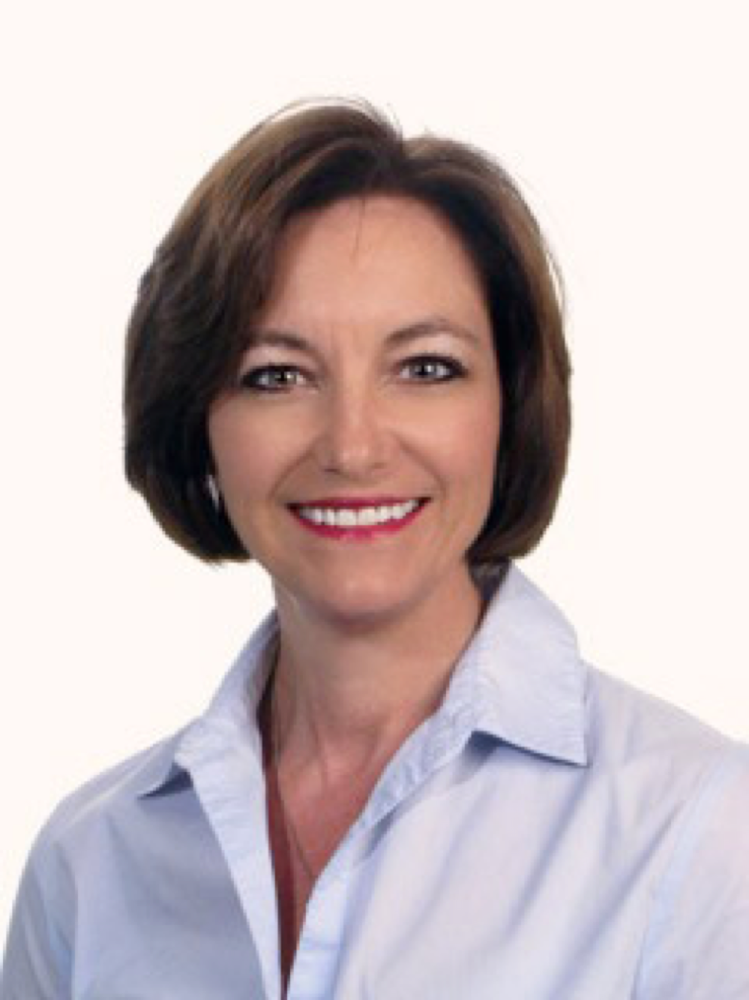}}] {Carrol Plummer} is the CEO and a co-founder of Vivent SARL, a research intensive Swiss based business focused on understanding communications networks in biological systems. She received her BSc in Mechanical Engineering from University of Calgary in 1981, and an MBA from INSEAD, France in 1990. Her research interests include mechanical and electrical signaling in plants, bacteria and animals from both theoretical and applied perspectives, and the role signaling plays in organism wide coordinated responses such as growth or defense reactions.  She is currently working on research supported by grants from DARPA and Innosuisse, with a focus on the use of machine learning to analyse electrophysiology signals.
\end{IEEEbiography}
\vskip 0pt plus -1fil
\begin{IEEEbiography}[{\includegraphics[width=1in,height=1.25in,clip,keepaspectratio]{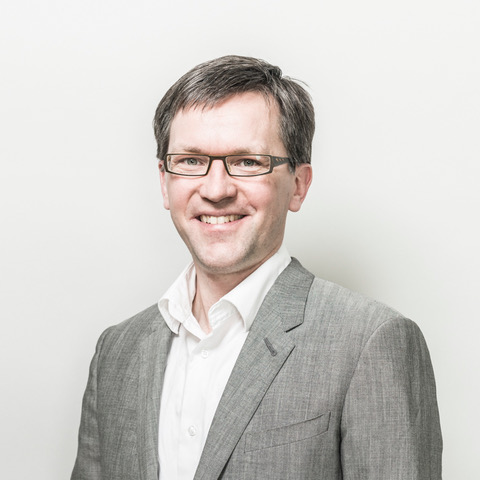}}] {Andrew Eckford} (M’96–S’99–M’03–SM’15)  is an Associate Professor in the Department of Electrical Engineering and Computer Science at York University, Toronto, Ontario. He received the B.Eng. degree from the Royal Military College of Canada in 1996, and the M.A.Sc. and Ph.D. degrees from the University of Toronto in 1999 and 2004, respectively, all in Electrical Engineering. Andrew held postdoctoral fellowships at the University of Notre Dame and the University of Toronto, prior to taking up a faculty position at York in 2006. Andrew’s research interests include the application of information theory to nonconventional channels and systems, especially the use of molecular and biological means to communicate. Andrew’s research has been covered in media including The Economist and The Wall Street Journal, and was a finalist for the 2014 Bell Labs Prize. Andrew is also a co-author of the textbook Molecular Communication, published by Cambridge University Press.
\end{IEEEbiography}
\vskip 0pt plus -1fil

\end{document}